\documentclass[showpacs,preprintnumbers]{revtex4-2}
\usepackage{amssymb}
\usepackage{graphicx}
\usepackage{color}
\usepackage{ulem}

\def\bc{\begin{center}}
\def\ec{\end{center}}

\def\beq{\begin{equation}}
\def\eeq{\end{equation}}

\usepackage{graphicx}
\usepackage{caption}
\usepackage{subcaption}
\usepackage[T1]{fontenc}
\usepackage[utf8]{inputenc}
\usepackage[english]{babel}
\usepackage[hidelinks]{hyperref}
\usepackage{indentfirst}
\usepackage{amssymb}
\usepackage{amsmath}
\usepackage{mathrsfs}
\usepackage{braket}
\usepackage{float}
\usepackage{color}

\begin{document}

\title{Polaritonic and Excitonic Time Crystals based on TMDC strips in an external periodic potential}
\author{Gabriel P. Martins$^{1,2,3}$, Oleg L. Berman$^{1,2}$ and Godfrey Gumbs$^{2,3,4}$}

\affiliation{$^{1}$Physics Department, New York City College
of Technology,          The City University of New York, \\
300 Jay Street,  Brooklyn, NY 11201, USA \\
 $^{2}$The Graduate School and University Center, The
City University of New York, \\
365 Fifth Avenue,  New York, NY 10016, USA\\
 $^{3}$Department of Physics and Astronomy, Hunter College of The City University of New York, 695 Park Avenue, \\
New York, NY 10065, USA\\
$^{4}$Donostia International Physics Center (DIPC), P de Manuel Lardizabal, 4, 20018 San Sebastian, Basque Country, Spain }
\date{\today}

\begin{abstract}

{We investigated the dynamics of  {Bose-Einstein condensates (BECs) under an external periodic potential. We consider two such systems, the first being made} of exciton-polaritons in a nanoribbon of transition metal dichalcogenides (TMDCs), such as MoSe$_2$, embedded in a microcavity { with a special curvature, which serves as the source of the external potential. The second, made of bare excitons in a nanoribbon of twisted TMDC bilayer, which naturally creates a periodic Moiré potential that can be controlled by the angle of twist. We }  proved that such systems exhibit a Time Crystal (TC) phase.  This was demonstrated by the fact that the calculated {BEC} spatial density  profile shows a non-trivial two-point correlator that oscillates in time.  These {BECs} density profiles were calculated by solving the quantum Lindblad master equations for the  density matrix within the mean-field approximation. We then go beyond the usual mean-field approach, by adding a stochastic term to the master equation, which corresponds to quantum corrections, and we show that the TC phase is still present.}


\end{abstract}

\maketitle

\section{Introduction}
\label{intro}

\medskip
\par
Symmetry breaking is a subject of much interest in condensed matter physics, since it leads to further understanding of many phases of matter exhibited by materials. In symmetry-broken phases, a system has a Hamiltonian which is unchanged by a particular symmetry transformation, but the state of the system itself does not exhibit the same symmetry. The breaking of translational symmetry, for example, leads to a crystal phase; the breaking of spin-rotational symmetry gives rise to a ferromagnetic phase. A time crystal (TC) is a state of matter that shows spontaneous breaking of time translation symmetry (TTS). This phase was first predicted by Wilczek in 2012 \cite{Wilczek} and has been the subject of many studies ever since. {In their recent paper, Watanabe and Oshikawa provided a proper mathematical definition for a TC by following steps based on the mathematical definition of a spatial crystal \cite{Watanabe}.} Many no-go theorems have been advanced showing that TCs cannot exist under certain conditions \cite{Watanabe,Noz, Bruno}. Ever since then, several groups have proposed different settings that could exhibit a TC phase \cite{Khemani,  Else, Keyser}. {In a recent review, Sondhi \textit{et al.} presented a detailed and very well written description of TCs \cite{Sondhi}. 
Systems ranging from chains of spins \cite{spin} to   BECs in many different settings \cite{Kavokin,liao,wright,wang} have been shown to present spontaneous breaking of the TTS. It has been shown that a chain of spins with long-range interactions can form a robust TC at zero temperature \cite{kozin}.}   Dissipative TC in an asymmetric nonlinear photonic dimer was proposed in Ref.~\cite{Savona}.  It has recently been theorized that a set of  TCs could be used as a model to simulate the human brain \cite{brain}.

\medskip
\par

{
Absorption of a photon by a semiconductor leads to  the creation of an electron in the conduction band and a positive charge, i.e., ``hole'', in the valence band. This electron-hole pair can form a bound state called  an ``exciton'' {\cite{Moskalenko_Snoke}}.  The Bose-Einstein condensation  and superfluidity  of such excitons are  expected to exist   at experimentally  observed exciton densities at temperatures much higher than for the BEC of alkali atoms \cite{Moskalenko_Snoke,excond}. A direct exciton is a two-dimensional (2D) exciton, formed as a bound state by an electron and a hole in a single semiconductor quantum well while an indirect exciton is formed by the bound state of an electron and a hole in neighboring quantum wells. Excitons can be created when the material absorbs photons and can decay by emitting photons. When a suitable material for the occurrence of excitons is put inside an optical microcavity, linear superposition between photons and excitons can be found {\cite{Snoke_Keeling}}. Such a quasi-particle is known as an exciton-polariton.}

\medskip
\par
Many theoretical and experimental studies have  identified Bose coherent effects of 2D excitonic polaritons in a quantum well embedded in a semiconductor   microcavity~\cite{Littlewood,Carusotto_rmp,Snoke_book,Snoke_Keeling}. To obtain polaritons, two Bragg mirrors are placed opposite {to} each other in order to form a microcavity, and a quantum well is embedded within the cavity at the antinodes of the confined optical mode. The resonant interaction between a direct exciton in a quantum well and a microcavity photon results in the Rabi splitting of the excitation spectrum. Two polariton branches appear in the spectrum due to the resonant exciton-photon coupling. The lower polariton branch of the spectrum has a minimum at zero momentum. The effective mass of the lower polariton is extremely small. These lower polaritons form a 2D weakly interacting Bose gas. The extremely light mass of these bosonic quasiparticles at experimentally achievable excitonic densities results in a relatively high critical temperature for superfluidity. The critical temperature is relatively high because the 2D thermal de Broglie wavelength is inversely proportional to the mass of the quasiparticle, and this wavelength becomes comparable to the distance between the bosons. {BEC and superfluidity of exciton-polaritons have been observed in a microcavity~\cite{Carusotto_rmp,Snoke_book,pol_fluid}}. The various applications of microcavity polaritons for optoelectronics and nanophotonics have  been developed recently~\cite{Snoke_Keeling}.

\medskip
\par

Two-dimensional van der Waals materials such as the atomically thin transition metal dichalcogenides (TMDCs) have unique physical properties, which are attractive for a broad range of applications.  Monolayers of TMDC such as $\mathrm{Mo S_{2}}$, $\mathrm{Mo Se_{2}}$, $\mathrm{Mo Te_{2}}$, $\mathrm{W S_{2}}$,  $\mathrm{W Se_{2}}$, and $\mathrm{W Te_{2}}${, for instance}, are 2D  direct bandgap semiconductors, which have a variety of applications in electronics and optoelectronics~\cite{Kormanyos}. {The strong interest in TMDC monolayers is  motivated by the following factors: 
the direct gap in the band structure spectrum~\cite{Mak2010}, the existence of excitonic valley physics, and the possibility of electrically tunable, strong light-matter interactions~\cite{Xiao,Mak2013}.  
} Monolayer TMDCs { hold great appeal for electronics and} have already been implemented in field-effect transistors, logic devices, and lateral and tunneling optoelectronic structures~\cite{Kormanyos}. { The  monolayer TMDCs have hexagonal lattice structures, and the nodes (valleys) in the dispersion relations of the valence (conduction) band can be found at the $\mathbf{K}$ and $\mathbf{K}^{\prime}$ points of the hexagonal Brillouin zone.} It is important that these 2D crystals do not have inversion symmetry~\cite{Kormanyos}. The specific properties of excitons in monolayer TMDCs have been the subject of many experimental and  theoretical studies { (see, for example,} \cite{Glazov_rmp}).  The large binding energy and long lifetime of interlayer excitons in van der Waals heterostructures have prompted much work on these materials~\cite{Rivera,Latini}.   The exciton-polaritons, formed by direct excitons  in a TMDC monolayer embedded in a microcavity and  cavity photons were observed experimentally at room temperature~\cite{Menon}. { The superfluidity of exciton-polaritons in a TMDC monolayer
 embedded in a microcavity has been studied in Refs. \cite{BLS,BKL,BKLS}.}

\medskip
\par

Stacking 2D materials { to form} van der Waals heterostructures opens up new strategies for materials properties engineering. One increasingly important example is the possibility of using the relative orientation (twist) angle between a pair of 2D crystals to tune electronic properties. For small twist angles and lattice constant mismatchings, heterostructures exhibit long period Moir\'{e} patterns, characterized by the periodic potential acting on the charge carriers. In 2D materials, a Moir\'{e} pattern with a superlattice potential can be formed by vertically stacking two layered materials with a twist and/or a difference in lattice constant. It is well established that the Moir\'{e} superlattice can modulate the electronic band structure of the material and lead to transport properties such as unconventional superconductivity~\cite{Cao} and insulating behavior driven by correlations~\cite{Hunt,Dean,Kim}.

\medskip
\par
In bilayer TMDC, formed by vertically stacking two TMDC  monolayers, intralayer excitons are formed by an electron and a hole, located in the same monolayer. On the other hand,  interlayer excitons are formed by an electron and a hole, located in two neighboring monolayers
. Since an  interlayer exciton is composed of an electron and a hole that are separated in neighboring layers, its properties can depend strongly on the layer configurations and external fields. For example, it was recently predicted that Moir\'{e} superlattices, where the interlayer atomic registry changes periodically over space, can host arrays of localized interlayer exciton states with distinct valley selection rules~\cite{Yu,Wu}. The Moir\'{e} degree of freedom for interlayer excitons offers exciting opportunities for realizing quantum emitter sources~\cite{Yu,Wu}. It was demonstrated that the optical  absorption spectrum of TMDC bilayers is drastically altered by long period Moir\'{e}   patterns that introduce twist-angle dependent  satellite excitonic peaks~\cite{Wu_PRL}.

\medskip
\par

The observation of multiple interlayer exciton resonances with either positive or negative circularly polarized irradiation was reported in a $\mathrm{Mo Se_{2}}$/$\mathrm{W Se_{2}}$ heterobilayer with a small twist angle~\cite{Tran}. These resonances were  attributed to excitonic ground and excited states confined within  the Moir\'{e} periodic potential. This interpretation is supported by recombination dynamics and by the dependence of these interlayer exciton resonances on twist angle and temperature~\cite{Tran}. The experimental evidence of interlayer valley excitons trapped in a  Moir\'{e} potential in $\mathrm{Mo Se_{2}}$/$\mathrm{W Se_{2}}$ heterobilayers was reported~\cite{Seyler}. At low temperatures, photoluminescence was observed close to the free interlayer exciton energy but with linewidths over one hundred times narrower (around $100 \ \mathrm{\mu eV}$). For the semiconducting heterostructures assembled from incommensurate $\mathrm{Mo Se_{2}}$ and $\mathrm{W S_{2}}$ monolayers, it was demonstrated   that excitonic bands can hybridize, resulting in a resonant enhancement of  Moir\'{e} superlattice effects~\cite{Alexeev}.  The observation of multiple interlayer exciton states coexisting in a $\mathrm{W Se_{2}}$/$\mathrm{W S_{2}}$ Moir\'{e} superlattice was reported~\cite{Jin}. Spin, valley and Moir\'{e} quasi-angular momentum of these multiple interlayer exciton states were determined through novel resonant optical pump-probe    spectroscopy and photoluminescence excitation spectroscopy~\cite{Jin}.

\medskip
\par

{ In this paper we study the dynamics of {two Bose-Einstein Condensates. One made of exciton-polaritons on a strip of MoSe$_2$, and the other made of bare excitons on a strip of twisted bilayer TMDC.} Under some circumstances, these systems can be shown to be in a Time Crystal phase. In the dilute regime, the systems will form a BEC whose mean-field dynamics is given by Gross-Pitaevskii's equation (GPE). { For the exciton-polaritons,} we consider the strip to be inside a spatially curved  optical microcavity. The curvature of the cavity creates an effective external potential on the photons inside the cavity, which, in turn, acts as an effective potential for the polariton as a whole \cite{BLS,BKL,BKLS}. This effective potential enters directly in the GPE and significantly changes the dynamics of the condensate. We calculate the curvature required for an effective sinusoidal potential on the polariton BEC. { For the bare excitons system, the Moiré pattern caused by the relative twist between the two TMDC layers creates a {periodic} external potential for the excitons in the strip. We then present the mathematical condition  for a system to be in a TC phase \cite{Watanabe}. After that, we properly present the GPEs that govern the mean-field dynamics. We then explain how to go beyond the usual mean-field analysis by adding a stochastic term corresponding to inherent quantum uncertainty to the GPE \cite{Snoke_book}. } We show some computational results of the time-evolution of { both} BECs' mean-field given by the modified GPE we considered, where we see clear evidence of the Time-Translation Symmetry breaking that characterizes TCs. { For both of our considered systems, we provided the proper mathematical proof that they are pertinent, Time Crystals.  We do this by following a procedure similar to the one done in  Ref.\  \cite{kozin}, in which authors demonstrated that a chain of spins with long-range interactions present a TC phase, by showing the two-point correlator in the total magnetization of the system, analyzed at distant times, exhibits some non-trivial time-dependence. Similarly, we analyze the two-point correlator of the condensate density at two distant positions on the strip and a long time apart, and also show that such a correlator exhibits non-trivial time dependence.} This rigorous mathematical proof has not been done by other propositions of BEC-based TCs in the literature \cite{Kavokin,liao,wright}. { We also show that our results still hold even when} we go beyond the usual mean-field description by adding quantum corrections to the GPE.} 


\medskip
\par
The rest of the paper is organized as follows. { We present the theoretical foundation of the paper in Sec. \ref{theory}, in which we define both systems we studied alongside presenting the mathematical definition of a Time Crystal. We start this section by considering a BEC of exciton-polaritons in a strip of TMDC, and we show how an uneven microcavity leads to an effective external potential to polaritons. After that, we explain how a BEC of bare excitons in a twisted TMDC bilayer experiences a similar external potential. In Sec. \ref{math}, we present model  equations for the dynamics of a non-equilibrium BEC in a periodic potential has been presented, both in the usual mean-field description, as well as showing how to go beyond that, by adding a stochastic term corresponding to quantum uncertainty in the equations. Lastly, we show our numerical results of our calculations for the dynamics of an exciton-polariton and bare-exciton condensate in a periodic potential are presented in Sec.~\ref{numeric}. There we present definite proof that both systems behave as TCs, even when quantum uncertainty is taken into consideration.}  We summarize the results of our work in  Sec.~\ref{conc}.

{
\section{Theoretical Framework}\label{theory}

In this section, we provide the theoretical framework for the remainder of the paper. First, we define our systems of interest, namely a polariton condensate on a strip of TMDC inside a spatially curved microcavity and an excitonic BEC on a strip of twisted TMDC bilayer. After that, we provide the mathematical definition of a Time Crystal as proposed by Watanabe and Oshikawa \cite{Watanabe}.

\subsection{Exciton-Polariton BEC in an Uneven Microcavity}
}

\begin{figure}[h]
\begin{center}
\includegraphics[width = 0.5\textwidth]{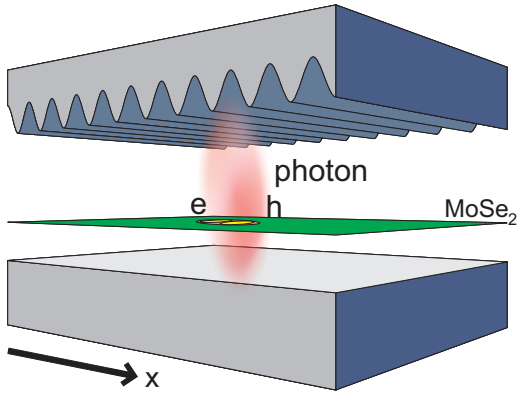}
\end{center}
\caption{ Schematic  representation  of the considered system. A strip of MoSe$_2$ inside an uneven microcavity. The cavity is composed of a plane mirror at the bottom and a spatially curved mirror on the top. The cavity length is, therefore, not constant and is a function of $x$, $L_C(x)$, given by Eq. (\ref{len}).}
\label{schem}
\end{figure}

{ In semiconductors, electrons in the conduction band and holes in the valence band can be bound together, forming an hydrogen-like structure called an exciton \cite{Moskalenko_Snoke}. The creation energy for the excitons $\varepsilon_{exc}$ is equal to the gap energy between the conduction and valence bands of the semiconductor, $\Delta$, minus the binding energy of the electron-hole pair $\varepsilon_b$, $\varepsilon_{exc} = \Delta -\varepsilon_{b}$. Excitons are created when the material absorbs photons and can spontaneously decay by the recombination of the electron-hole pair, which, in turn, emits a photon. When a material that has excitons is contained within an optical microcavity, in which photons are confined, a superposition state can be formed between excitons and photons. Those states are called exciton-polaritons \cite{Snoke_Keeling}. There are two branches of exciton-polaritons, one with higher energy, called the \textit{upper polariton} branch, and one with lower energy, called the \textit{lower polariton} branch. { Since they behave as bosonic quasiparticles, { lower polaritons can form Bose-Einstein condensates } \cite{Snoke_Keeling}. One such condensate has been recently verified in room-temperature settings \cite{pol_fluid}.}}

{ Inside an optical microcavity of length $L_C$, the energy of a photon  of the $q$-th mode with momentum $\mathbf{P}$ is $\varepsilon_{ph}(\mathbf{P}) = (c/n)\sqrt{P^2+\hbar^2\pi^2{L_C}^{-2}}$, where $n = \sqrt{\epsilon_r}$ is the refractive index of the cavity. For low momentum, $\varepsilon_{ph} \approx \dfrac{\hbar\pi q c}{n L_C}$. If this cavity is uneven, namely, if one of the mirrors that make it is curved, the energies of the photons depend on the position. This effectively creates an external potential, $V_{\mathrm{ph}}$, for low-momentum photons inside the cavity. If, for example, a cavity has a length that varies in the $x$ direction like
\begin{equation}
L_C (x) = \frac{\hbar\pi q c}{n\left(\varepsilon_{0}+V_{\mathrm{ph}}(x)\right)},\label{len}
\end{equation}
low-momentum photons of the $q$-th mode in such a cavity will have an energy  $\varepsilon_{\mathrm{ph}}(x) = \varepsilon_0 +V_{\mathrm{ph}}(x)$. It is evident that the effect of such a curvature is equivalent to the addition of an external potential to the photons. It can be shown that an effective potential on either photons, $V_{\mathrm{ph}}(x)$, or excitons, $V_{\mathrm{exc}}(x)$, leads to an effective potential on polaritons, which can be approximated as $V_{\mathrm{eff}}(x)=\dfrac{1}{2}\left(V_{\mathrm{ph}}(x)+V_{\mathrm{exc}}(x)\right)$ \cite{BLS,BKL,BKLS}.}

\medskip
\par

{Our goal is to study the effect of an external periodic potential on a BEC of exciton-polaritons in a strip of TMDC. According to Eq. (\ref{len}), this can be achieved by a cavity with length $L_C(x)$ given by
\begin{equation} L_C (x) = \frac{\hbar\pi q c}{n\left(\varepsilon_{\mathrm{exc}}+2V_0 \cos(kx)\right)}  \label{len2}  .
\end{equation}
Polaritons confined in such a cavity would be subjected to an effective potential $V_{\mathrm{eff}}(x) = V_0\cos (kx)$.
}

\subsection{Excitonic BEC in a Twisted TMDC Bilayer}

\begin{figure} [H]
\begin{center}
\includegraphics[width=0.8\textwidth]{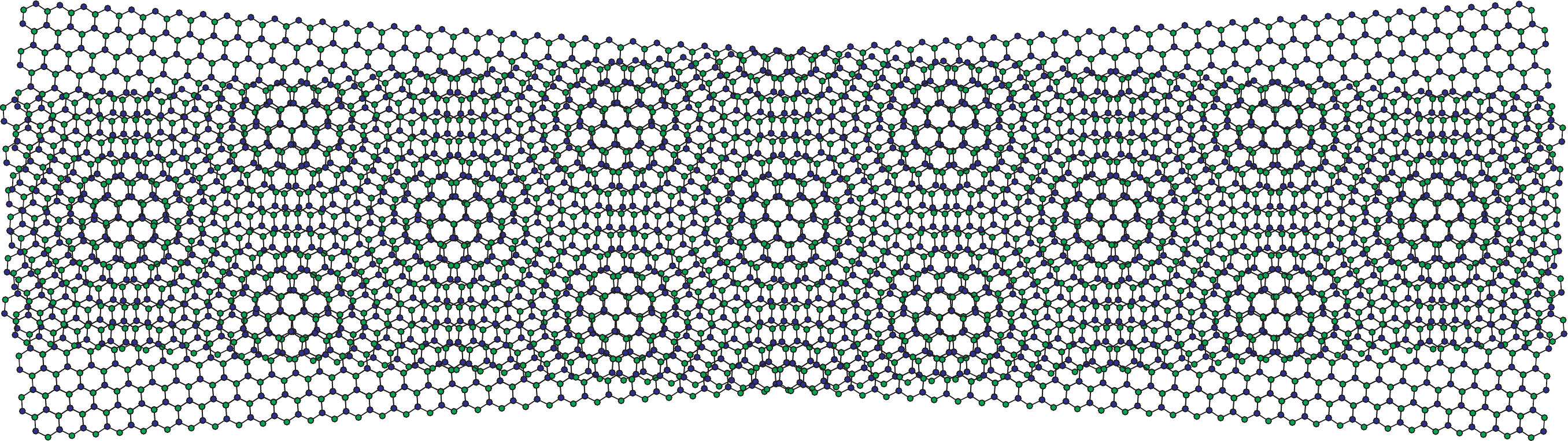}
\end{center}
\caption{{ Moiré pattern  in the crystal lattice structure seen by twisting one of the layers of a bilayer TMDC.  The pattern is created by the difference in atomic alignment in the upper and lower layers. Excitons in such an environment will be subject to an external periodic potential given by Eq. (\ref{moire}).  \label{fmoire} }}
\end{figure}

{ Throughout this paper, we will also consider a system composed of bare excitons in a twisted TMDC bilayer. The superposition of two layers of TMDC twisted with respect to each other  will act on the excitons in it as an external periodic potential \cite{Tran}. Figure\    \ref{fmoire} depicts a twisted TMDC bilayer, where we can see the Moiré pattern that is formed. }

Bare excitons, just like exciton-polaritons, can form a BEC phase {at low temperatures}.  When excitons are  in a twisted TMDC bilayer, it has been shown that they are subjected to an effective periodic potential $V(\mathbf{r})$ caused by the {Moir\'e} pattern given by \cite{Tran}
\begin{eqnarray}
\label{moire}
V(\mathbf{r}) &=& V_0\sum_j \cos\left(k\mathbf{b}_j\cdot\mathbf{r}\right) \nonumber \\
&=& 2V_0 \left( \cos (kx) + \cos k\left(\frac{x}{2}+\dfrac{\sqrt{3}}{2}y\right)+\cos k\left(\frac{x}{2}-\dfrac{\sqrt{3}}{2}y\right)\right)
\end{eqnarray}
where   $\mathbf{b}_j$ are unit vectors which divide the plane into six identical sections, { and $k=\frac{2\pi}{a_M}$, where $a_M$ is the Moiré period.} As in Ref. \cite{Tran}, we consider TMDCs with {an} hexagonal lattice. If we constrain the excitons to a ribbon of TMDC that is finite in the $\textit{y}$ direction with a width $W$ and infinitely long in the $\textit{x}$ direction, we can replace the potential by an effective potential $V_{\mathrm{eff}}(x)$ given by
\begin{equation}
V_{\mathrm{eff}}(x) = \frac{1}{W}\int_{-W/2}^{W/2}dy\  V(x,y) = 2V_0 \cos (kx),\label{vexc}
\end{equation}
only if the width {$ W = \dfrac{4\pi m}{k\sqrt{3}}=\dfrac{2m}{\sqrt{3}}a_M,$} for positive integer $m$'s. {With both of our studied systems defined, we will now present the proper mathematical definition of what a Time Crystal is.}  

\subsection{Definition of a Time Crystal}
\label{TC}

{ Here we present the proper mathematical definition of a Time Crystal (TC), a phase of matter that has been proposed in 2012 by Wilczek \cite{Wilczek} and has been the object of much interest since then.

As many other phases of matter, TCs arise when the system spontaneously breaks one of the symmetries of it's Hamiltonian. When a system spontaneously breaks, for example, translational symmetry, it becomes a crystal; if it breaks spin rotational symmetry, it becomes a magnet. TCs, on the other hand, appear when a system breaks the Time Translational Symmetry (TTS).}

{In 2015 Watanabe and Oshikawa provided a mathematical criterion to verify whether a systemm is in a TC phase  \cite{Watanabe,Sondhi}. They followed steps based on the mathematical definition of a spatial crystal, in which, for the system to be considered a crystal, it has to show a non-trivial two-point correlator in an observable $\hat{O}$ in space far apart $\left(\lim_{|\mathbf{r}-\mathbf{r}^\prime|\rightarrow\infty}\langle\hat{O}(\mathbf{r})\hat{O}(\mathbf{r}^\prime)\rangle = f(\mathbf{r}-\mathbf{r}^\prime)\right)$. Analogously, for a system to be a TC, it has to show some non-trivial two-point correlator in some observable $\hat{O}$ at two different and long-apart times, namely, }
{
\begin{equation}
\lim_{|\mathbf{r}-\mathbf{r}^\prime|\rightarrow \infty} \lim_{t-t^\prime\rightarrow \infty}\left\langle \hat{O}(\mathbf{r},t) \hat{O}(\mathbf{r}^\prime,t^\prime) \right\rangle = c(t), \label{cond}
\end{equation}}
for some non-trivial{, or, in other words, non-stationary,} function $c(t)$. {In this paper, we will be considering systems confined in strips that are suffiently narrow in the $y$ direction for them to be trated as one-dimensional, so we can replace $\mathbf{r}$ and $\mathbf{r}^\prime$ in Eq. \ref{cond}, by $x$ and $x^\prime$. The expectation value seen in Eq. (\ref{cond}) is to be taken over the entire sample, while maintaining $x-x^\prime$ constant, which leads to
\begin{equation}
c(t) = \lim_{\Delta \rightarrow \infty} \lim_{t-t^\prime\rightarrow \infty}\dfrac{1}{L_x}\int_{-L_x/2}^{L_x/2} \hat{O}(x,t) \hat{O}(x+\Delta ,t^\prime) dx, \label{cond2}
\end{equation}
where $L_x$ is the length of the strip.}

{
\section{Mathematical Framework}
\label{math}

In this section, we present the mathematical framework we will use to study the dynamics of the condensate. First, we will present the equation that governs the evolution of the mean-field, which is a semi-classical and deterministic approach for the condensate. After that, we will present a way of going beyond the usual mean-field description by adding the effect of quantum uncertainty to the otherwise deterministic mean-field evolution.

\subsection{Mean-Field evolution}

Here, we will present the modified Gross-Pitaevskii equation (GPE) which governs the mean-field dynamics BECs. The same equation is valid for the BEC of exciton-polaritons, and for the BEC of bare excitons, both under an external potential, $V_{\mathrm{eff}} (r)$, caused by different reasons, as explained in the previous section.  Both quasiparticles are described by a Hamiltonian $\hat{H}$ for  a weakly interacting dilute Bose gas~\cite{BLS}:

\begin{eqnarray}
\label{Ham}
\hat{H} = \int d^{2} r\  \   \hat{\psi}^{\dagger} (r) \left(- \frac{\hbar^2\nabla^{2}}{2M_{p}} + V_{\mathrm{eff}} (r)\right)  \hat{\psi} (r) + \frac{U_{\mathrm{eff}}}{2} \hat{\psi}^{\dagger} (r) \hat{\psi}^{\dagger} (r)  \hat{\psi} (r)  \hat{\psi} (r)  ,
\end{eqnarray}
where $\hat{\psi}^{\dagger} (r)$ and   $\hat{\psi} (r)$ are creation and annihilation Bose operators for the polaritons or excitons, $M_{p}$ is the effective mass of the quasiparticle, $V_{\mathrm{eff}}(r)$ is the effective periodic potential acting on them. and $U_{\mathrm{eff}}$  is the Fourier image of the pair polariton-polariton (or exciton-exciton) repulsion potential at zero momentum. { For excitons, this repulsion is given by $U_{\mathrm{eff}}=U_{\mathrm{ex}}=\frac{3\hbar^2}{M_{\mathrm{ex}}}$, where $M_{ex}$ is the exciton mass; and $U_{\mathrm{eff}}=U_{\mathrm{pol}}=\frac{1}{4}U_{\mathrm{ex}}$, for polaritons \cite{BLS}.}}

\medskip
\par

The density matrix $\rho$ for non-equilibrium BEC can be obtained from  the quantum Lindblad master equation~\cite{Snoke_book}:

\begin{eqnarray}  \label{Lineq}
\frac{\partial \rho}{\partial t} = - \frac{i}{\hbar} \left[\hat{H},\rho \right] +  \int d^{2} r  \left(\kappa\mathcal{L}\left[\hat{\psi} (r) \rho\right] + \gamma\mathcal{L}\left[\hat{\psi}^{\dagger} (r),\rho\right] + \frac{\Gamma}{2}\mathcal{L}\left[\hat{\psi}^{2} (r),\rho\right] \right)   ,
\end{eqnarray}
where the usual Lindblad superoperator  is defined as

\begin{eqnarray}  \label{Linsup}
\mathcal{L}\left[\hat{X},\rho\right] = 2 \hat{X}\rho \hat{X}^{\dagger} - \left[\hat{X}^{\dagger} \hat{X}, \rho \right]_{+}    ,
\end{eqnarray}
and $\kappa$, $\gamma$ and $\Gamma$ are the rates { of} single-particle loss, single-particle incoherent pumping and two-particle loss, respectively. 

\medskip
\par

The mean-field equation of motion can be obtained by replacing {$\left\langle \hat{\psi} (r) \right\rangle = \tilde\varphi (r)$ (where $\tilde\varphi (r)$} is the wave function of the condensate) and decoupling all correlators. This procedure results in a modified Gross-Pitaevskii equation (GPE) \cite{Carusotto_rmp}, including dissipative terms describing particle gain and loss:
{
\begin{eqnarray}  \label{GPE1}
i \hbar\frac{\partial \tilde\varphi}{\partial t} = \left[- \frac{\hbar^2\nabla^{2}}{2M_{P}} + V_{\mathrm{eff}} (r) + U_{\mathrm{eff}}\left|\tilde\varphi\right|^{2} + i \left(\gamma - \kappa - \Gamma \left|\tilde\varphi\right|^{2}\right)\right] \tilde\varphi  .
\end{eqnarray}

{ When we consider our system to be a strip, with a width $L_y$ in the $y$ direction much smaller than the size of $L_x$ of the strip in the $x$ direction ($L_y\ll L_x$),} we can treat Eq. (\ref{GPE1}) as one-dimensional, by making the substitution $\tilde\varphi(r)\rightarrow\dfrac{\varphi(x)}{\sqrt{L_y}}$, which leads to
\begin{equation}\label{GPE}
i \hbar\frac{\partial \varphi}{\partial t} = \left[- \frac{\hbar^2}{2M_{P}}\dfrac{\partial^2}{\partial x^2} + V_{\mathrm{eff}} (x) +\dfrac{ U_{\mathrm{eff}}}{L_y}\left|\varphi\right|^{2} + i \left(\gamma - \kappa - \dfrac{\Gamma}{L_y} \left|\varphi\right|^{2}\right)\right] \varphi  .
\end{equation}
{ In our one-dimentional approximation for narrow strips, we have $V_{\mathrm{eff}} (x)$ sinusoidal for both systems{, namely, polariton BEC in a TMDC strip, embedded in a microcavity, and exciton BEC in a strip of a twisted TMDC bilayer,} for very different reasons. In the polaritonic system, this potential is created by the special manufacturing of the microcavity, while for the excitonic system, the potential arises naturally from the relative twist of the layers, as explained in the previous section.}


\subsection{Beyond mean-field description}\label{BMF}

Here, we explain what corrections should be added to Eq. (\ref{GPE}), in order for us to consider the effects of quantum randomness to the otherwise deterministic mean-field evolution.} 

In order to calculate the correlation and response functions, we will investigate the dynamics of polariton BEC beyond the mean-field description~\cite{Snoke_book}.   In the semiclassical limit, one can find the following equation for the  ``classical'' field  $\psi_{C}^{(ex)}$ {\cite{Snoke_book}}

\begin{eqnarray}  \label{feex}
i \hbar \frac{\partial \psi_{C}}{\partial t} = \left[-
\frac{\hbar^{2} \nabla^{2}}{2M} + V (\mathbf{r}) +
U\left|\psi_{C}\right|^{2} + i
\left(\gamma - \kappa-\Gamma|\psi_{C}|^2 \right)\right] \psi_{C} + i
\left(\kappa + \gamma \right)  \psi_{Q} ,
\end{eqnarray}
where $\psi_{Q}$  is the  ``quantum'' field, which can
be represented by the { dissipative-stochastic GPE (DSGPE). The DSGPE is} equivalent to
Eq.~(\ref{feex}) with the replacement $i \left(\kappa +
\gamma \right) \psi_{Q} \rightarrow \xi
(\mathbf{r},t)$, where $\xi (\mathbf{r},t)$ represents a
Gaussian white noise process with
\begin{eqnarray}  \label{corex}
\left\langle \xi (\mathbf{r},t) \right\rangle = 0, \hspace{1cm}
\left\langle \xi (\mathbf{r},t)  \bar{\xi}
(\mathbf{r}^{\prime},t^{\prime}) \right\rangle =
\frac{\left(\gamma + \kappa\right)}{2}\delta\left(t -
t^{\prime}\right)  \delta \left(\mathbf{r} -
\mathbf{r}^{\prime}\right).
\end{eqnarray}

\begin{figure}[H]
\begin{center}
\begin{subfigure}[b]{0.49\textwidth}
\includegraphics[width=\textwidth]{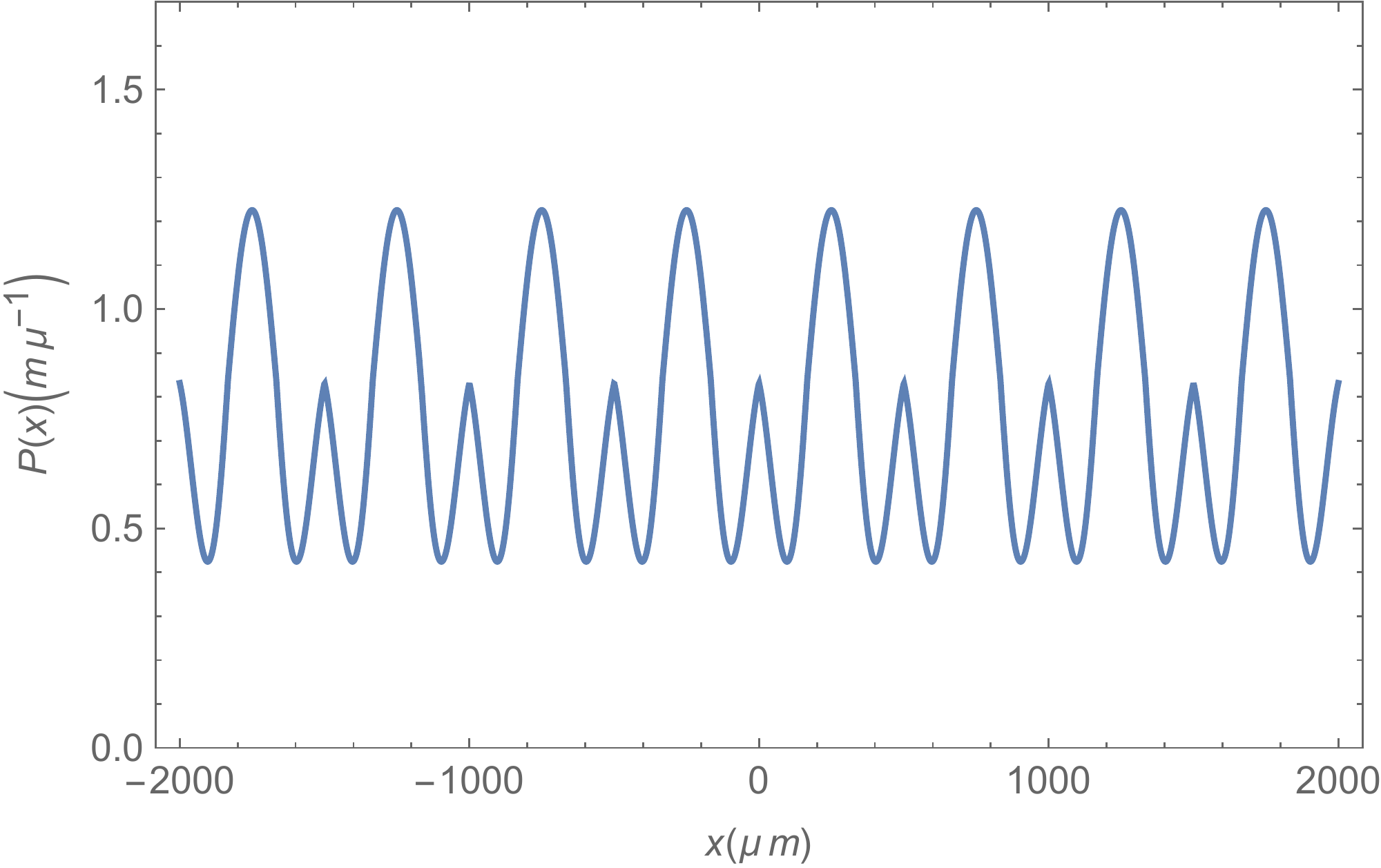}\\
\includegraphics[width=\textwidth]{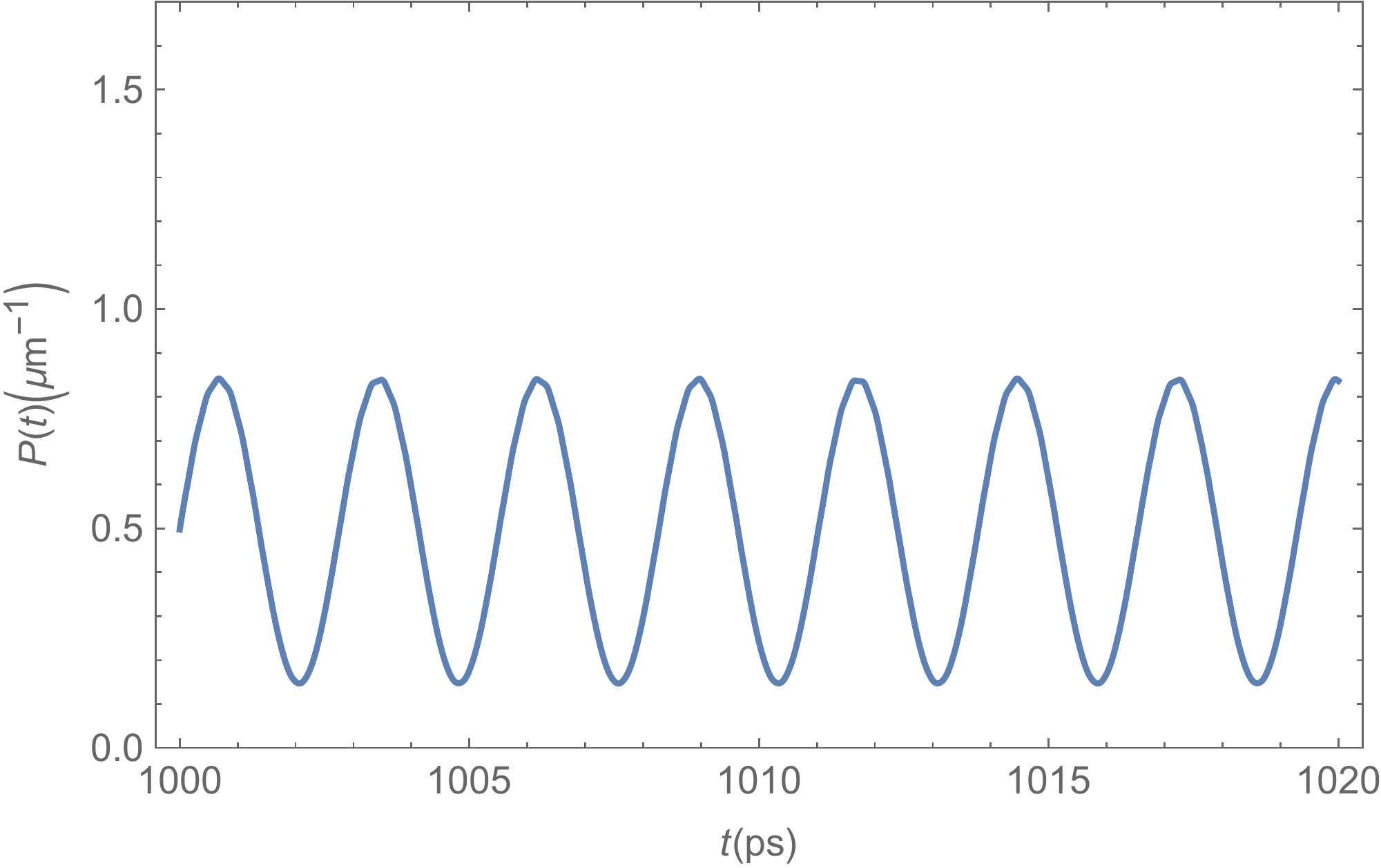}
\caption{}
\end{subfigure}
\hfill
\begin{subfigure}[b]{0.49\textwidth}
\includegraphics[width=\textwidth]{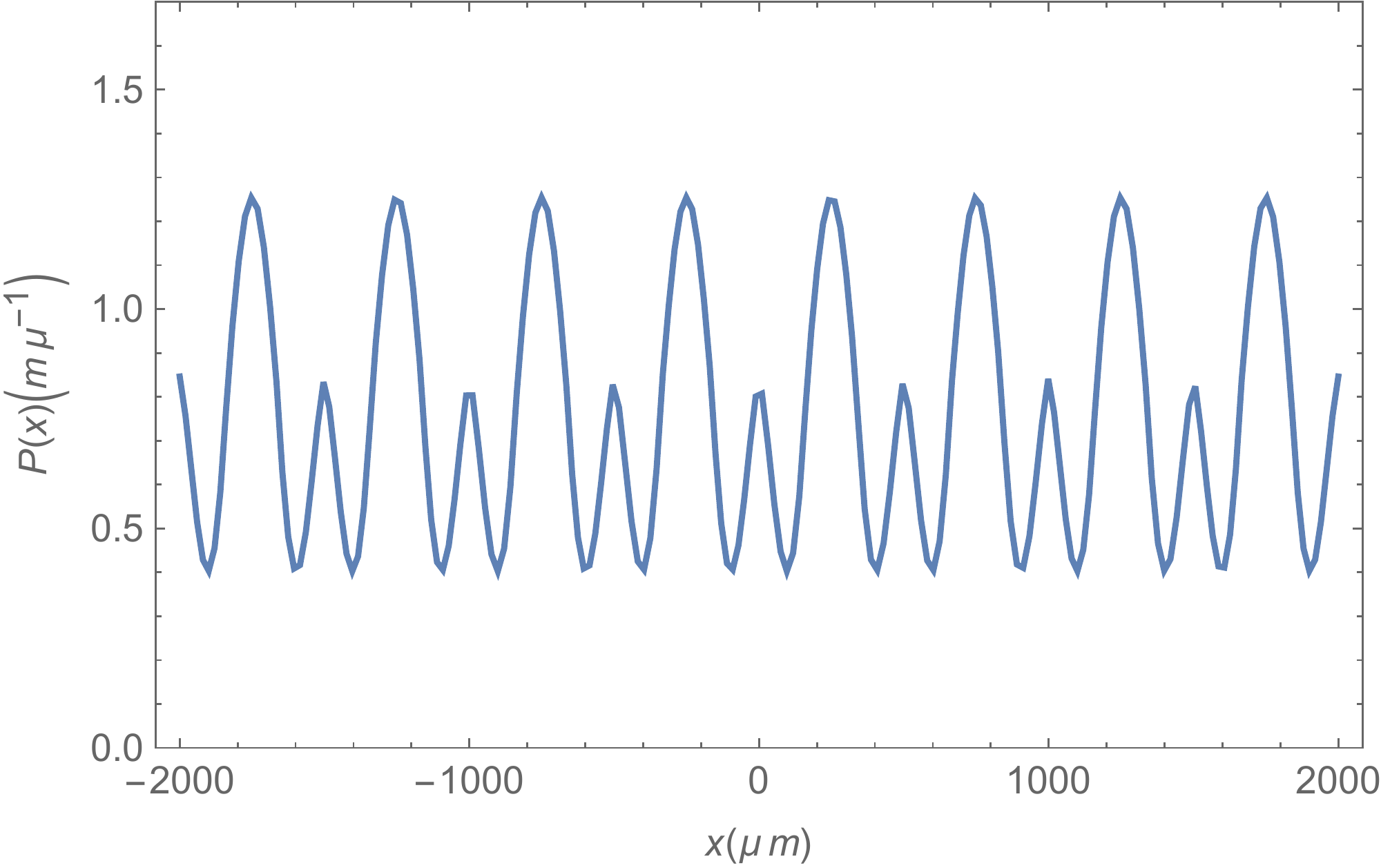}\\
\includegraphics[width=\textwidth]{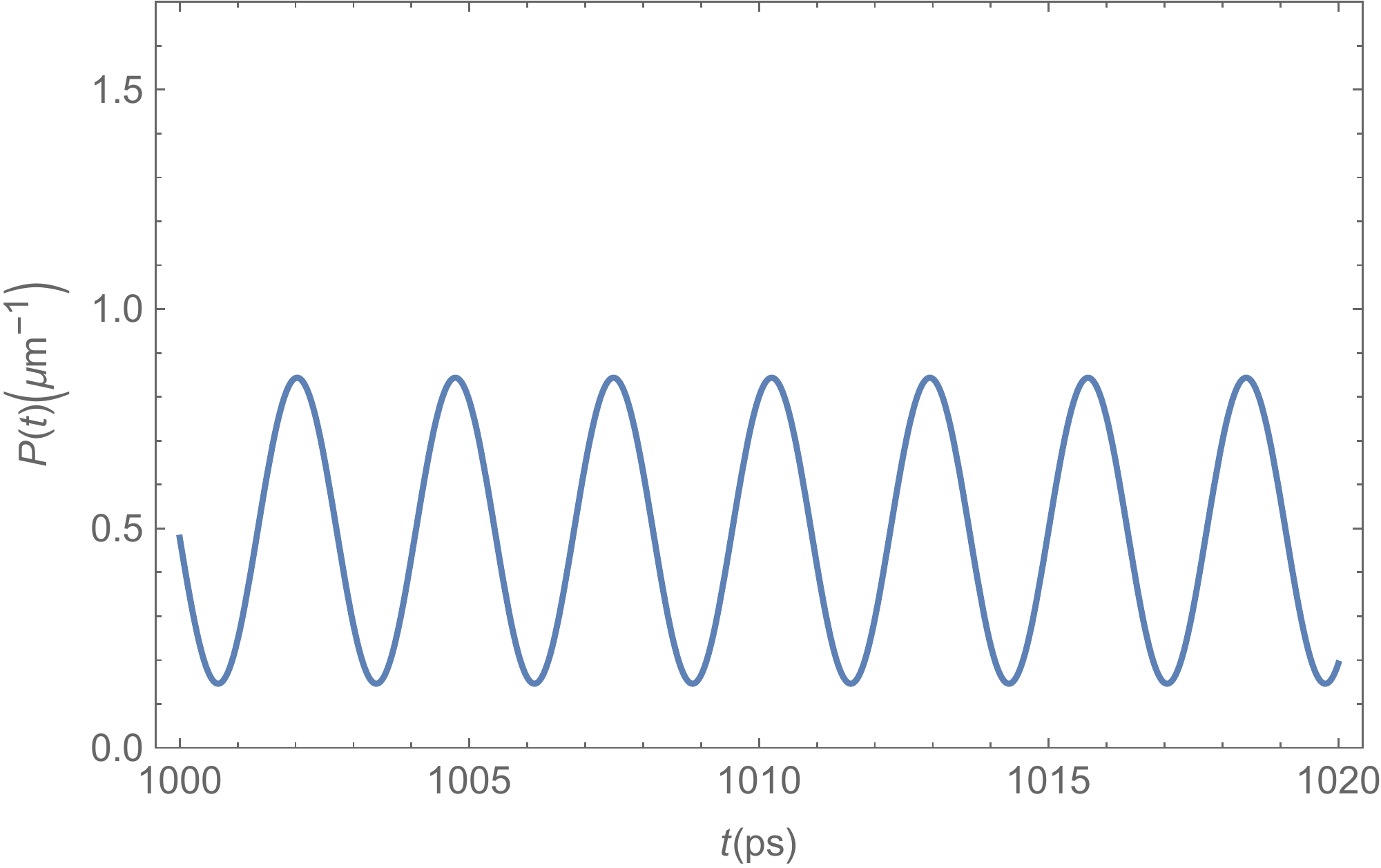}
\caption{}
\end{subfigure}
\end{center}
\caption{{ Comparison between the dynamics of the BEC { of exciton-polaritons} for two different initial conditions. On the top, we see the condensate distribution throughout the entire strip 1 ns after the beginning of the simulation; on the bottom, we see the time evolution for the condensate density at $x= 125$ $\mu$m for a period of 5 ps for the system starting at (a) the steady-state solution for the unperturbed condensate, {$\phi(x,t=0) = \sqrt{\dfrac{(\gamma-\kappa) L_y}{\Gamma}}$}; and (b) the vacuum state, {$\phi(x,t=0)=\epsilon$, with $\epsilon$ infinitesimal}. }\label{vac}}
\end{figure}

{For the one-dimensional case we are considering, the DSGPE becomes
\begin{equation}\label{DSGPE}
i \hbar\frac{\partial \psi_{C}}{\partial t} = \left[- \frac{\hbar^2}{2M_{P}}\dfrac{\partial^2}{\partial x^2} + V_{\mathrm{eff}} (x) +\dfrac{ U_{\mathrm{eff}}}{L_y}\left|\psi_{C}\right|^{2} + i \left(\gamma - \kappa - \dfrac{\Gamma}{L_y} \left|\psi_{C}\right|^{2}\right)\right] \psi_{C} +  \xi (x,t).
\end{equation}

\section{Numerical Results}\label{numeric}

In this section, we will present our numerical results. These results will be separated as such: first, we will do an in-depth analysis of the time evolution of the BEC of exciton-polaritons inside a spatially curved optical microcavity, where we will prove that such a system is in a TC phase both when we consider just the mean-field evolution, and when we go beyond the usual mean-field description. After that, we will analyze the evolution of the BEC of bare excitons on a twisted TMDC bilayer and show that the same results still apply.

\subsection{Polariton BEC on a spatially curved cavity}

First, we will consider a system of exciton-polaritons inside a spatially curved microcavity, as depicted in Fig.\  \ref{schem}. We consider the length of the cavity $L_C(x)$ to be given by Eq. (\ref{len2}).} {For our simulation, we consider a strip of MoSe$_2$, in which the polaritons have an effective mass $M_P = 5.8 \times 10^{-6} m_0$, where $m_0$ is the free electron rest mass. We consider this strip to be} inside a microcavity assembled with constituent elements similar to ones such as that in Ref.\   [\onlinecite{Cavity}], which has a refractive index $n=2.2$ and a mean width $L_c = 2.3$ $\mu$m. We consider the $q=5$ mode of that cavity, which resonates with the excitons in the MoSe$_2$ strip. The effective Rabi coupling between photons and excitons is $\hbar \Omega = 20.0$ meV. {This system has two characteristic lengths: the polariton's Bohr radius, $a_{2D}=\dfrac{\hbar^2\epsilon}{2M_P e^2}\approx 36$ $\mu$m, and the period of the effective {external} potential, {$a_{C}$, which we have chosen to be $a_C=$10 $\mu$m.} In order to avoid any edge effects, we need to consider the strip length to be much larger than these two values. Throughout all of our simulations, we considered the length of the strip to be $L_x = 4,000$ $\mu$m{, and its width to be $L_y= 1$ $\mu$m.

\medskip
\par

Our results showed that the polariton BEC density , $P(t,x) = |\varphi(t,x)|^2$, oscillates around the steady-state density, $P_0 = \dfrac{(\gamma-\kappa) L_y}{\Gamma}\approx 0.84$ $\mu$m$^{-1}$, for the case with no external potential. For almost all the plots the system was chosen to begin the simulation at $t=0$ in the steady-state of the unperturbed system, $\varphi(t=0,x)= \sqrt{P_0}$, the only exception being the ones in Fig. \ref{vac} (b). We will discuss this in further detail as we analyze each of the plots. }}

\begin{figure}[H]
\begin{center}
\begin{subfigure}[b]{0.32\textwidth}
\includegraphics[width=\textwidth]{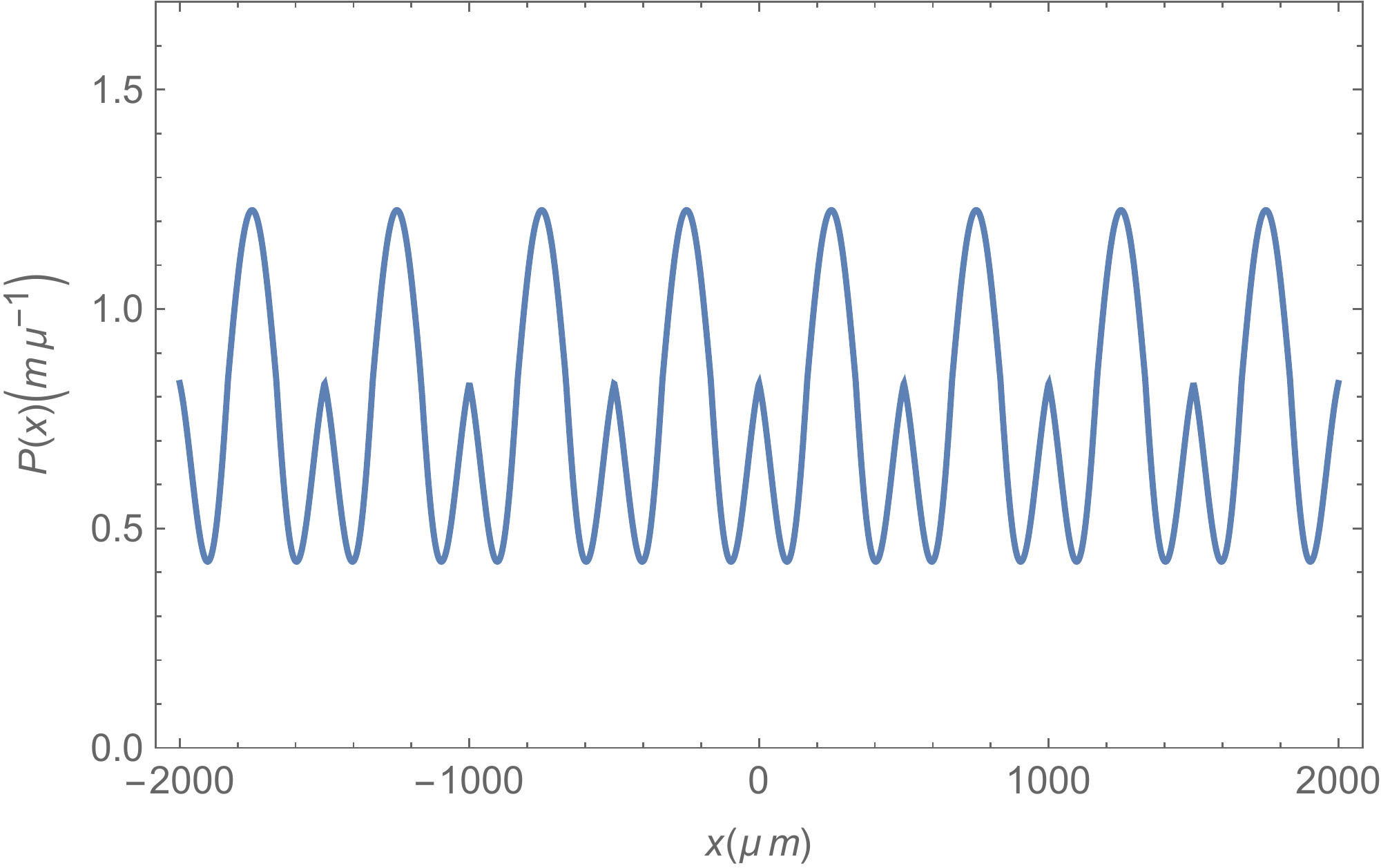}
\caption{}
\end{subfigure}
\hfill
\begin{subfigure}[b]{0.32\textwidth}
\includegraphics[width=\textwidth]{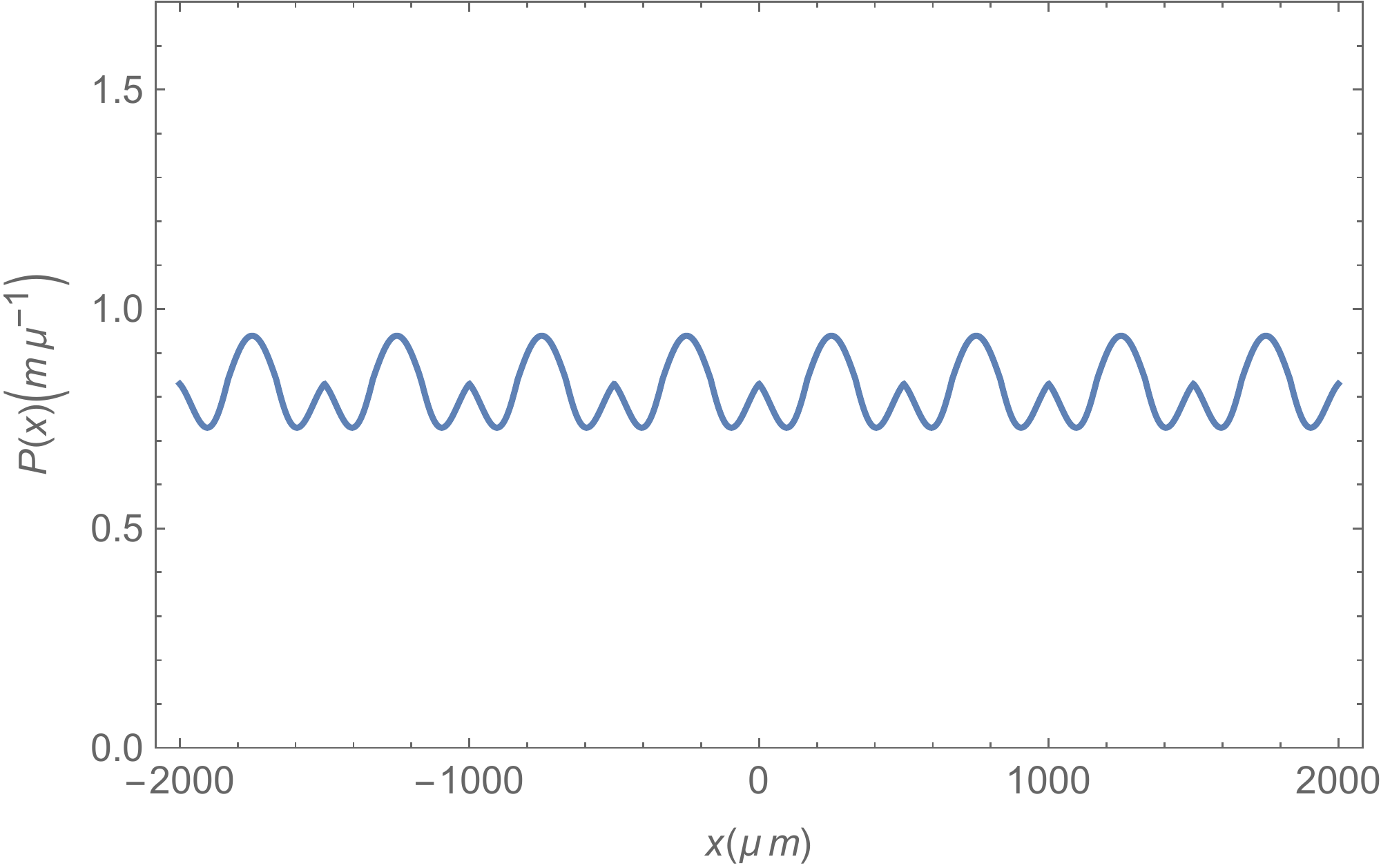}
\caption{}
\end{subfigure}
\hfill
\begin{subfigure}[b]{0.32\textwidth}
\includegraphics[width=\textwidth]{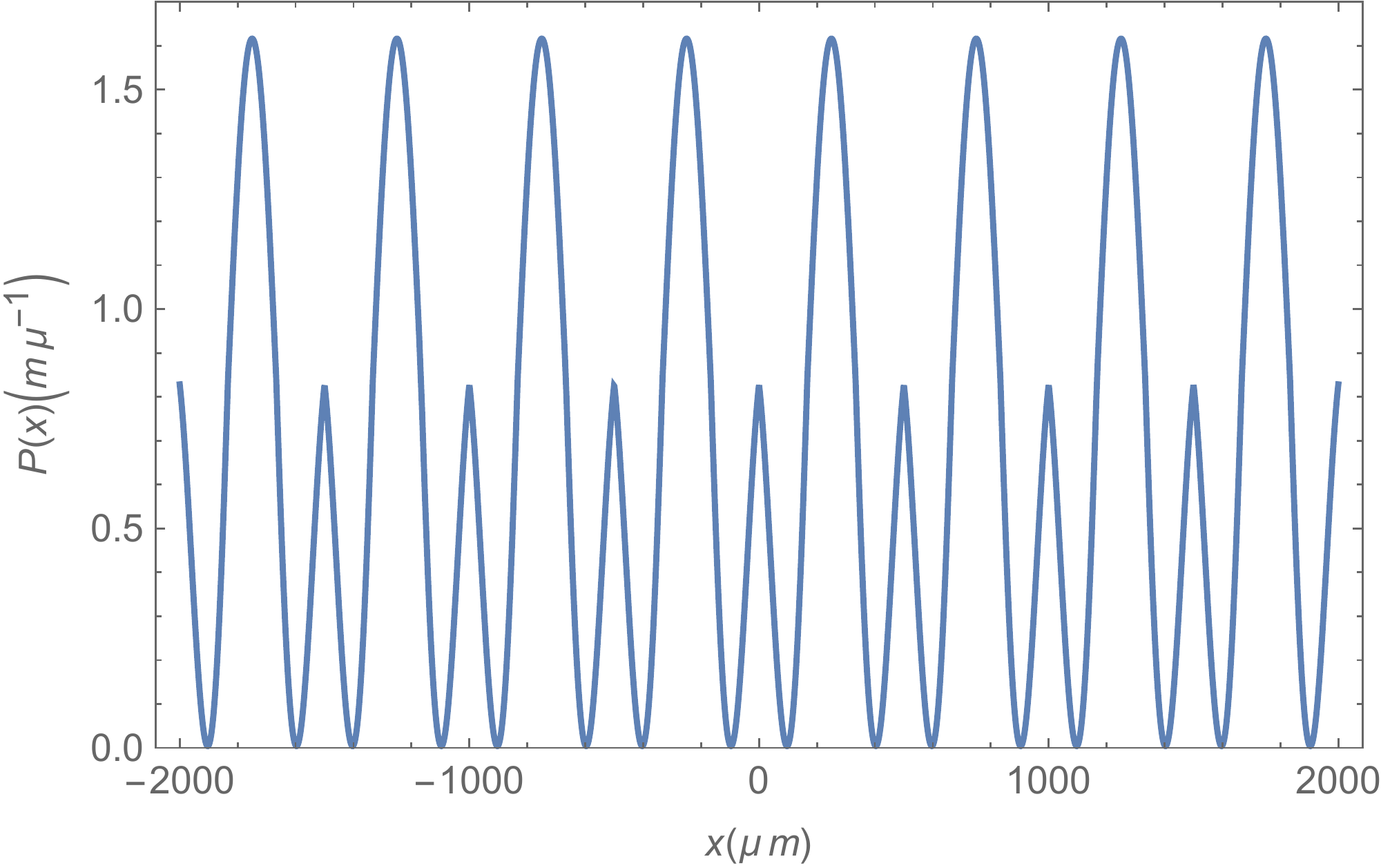}
\caption{}
\end{subfigure}
\end{center}
 \caption{{Polariton  BEC density $P(t,x) = |\varphi(t,x)|^2 $ for the polariton BEC in a ribbon of length 4000 $\mu$m at {three} different simulation times, at intervals of 1 ps from one another and starting at 3 ns.  (a) t = 3000 ps ; (b) t = 3001 ps; and (c) t = 3002 ps.}}
\label{times}
\end{figure}

{
\subsubsection{Mean-Field evolution}

We first restrained ourselves to the simpler case where we do not take quantum corrections into consideration. We did this by numerically solving Eq. (\ref{GPE}), and analyzing the results.      { Our first result depicts the polariton BEC density throughout the strip at {three} different times. { We allowed the simulation to run a long time before taking those plots in order to be sure that, if our system was able to thermalize and reach a steady state, it would have done so.} By looking at Fig. \ref{times}, we see that the polariton { BEC density throughout the strip is oscillating in time. We can see that this condensate appears to be pulsating around the steady state density for a planar microcavity, $P_0\approx 0.84$ $\mu$m$^{-1}$, since it deviates from this constant density a moderate amount at Fig. \ref{times} (a), then it approaches $P_0$ throughout the entire strip af Fig. \ref{times} (b) and deviates even more than on Fig. \ref{times} (a) on Fig. \ref{times} (c). Since the time difference between each of the plots in Fig. \ref{times} is of 1 ps, the period of these oscillations cannot be longer than just a few ps, meaning that our chosen start time of 3 ns is, indeed, sufficiently big to discard the possibility of the system thermalizing. The inability of reaching a steady state is one condition for the system to be a TC, which serves as a first evidence of this phase \cite{Sondhi}.}}} 


\medskip
\par


\medskip
\par

\begin{figure}[H]
\begin{center}
\hfill
\begin{subfigure}[b]{0.32\textwidth}
\includegraphics[width=\textwidth]{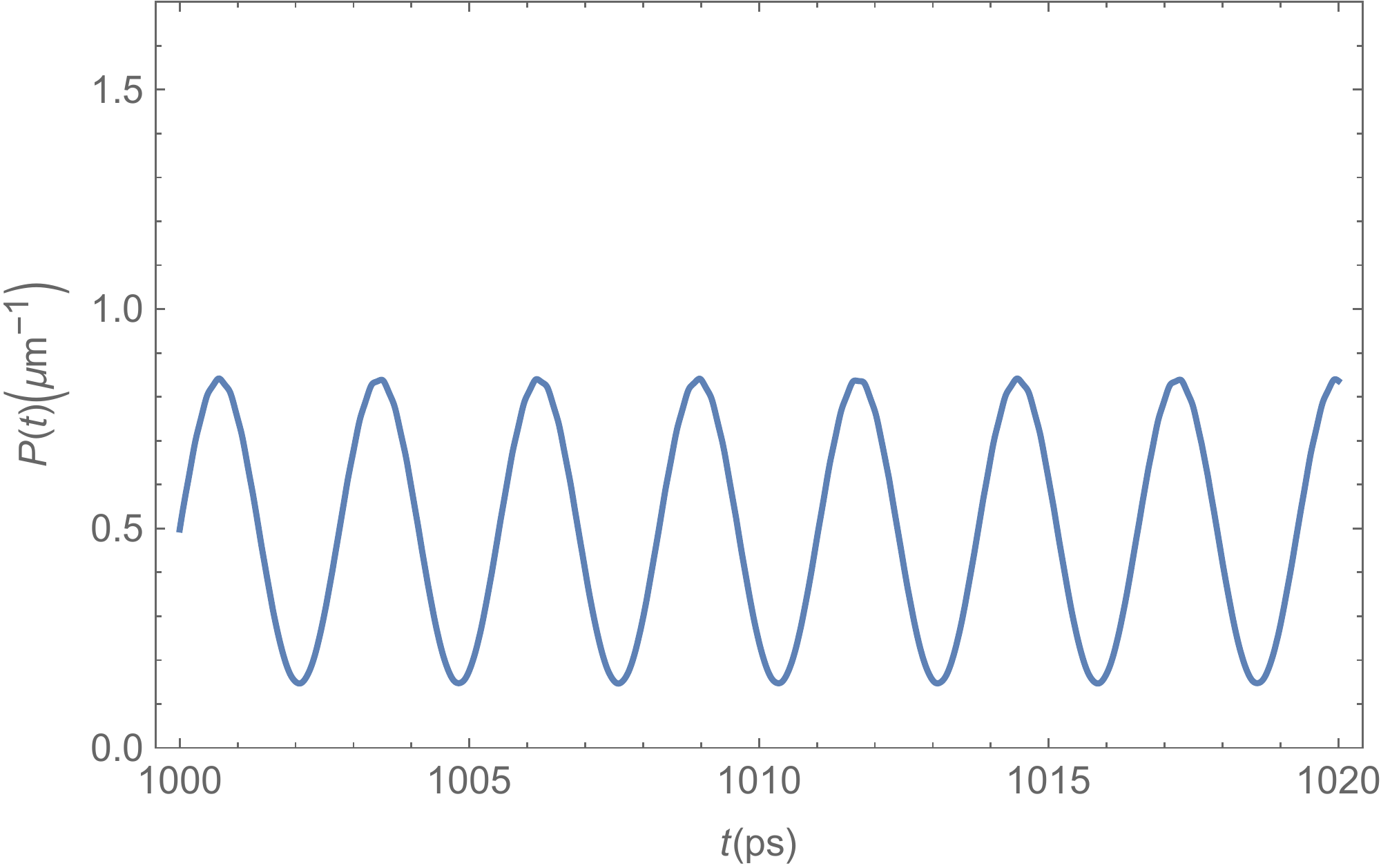}
\caption{}
\end{subfigure}
\hfill
\begin{subfigure}[b]{0.32\textwidth}
\includegraphics[width=\textwidth]{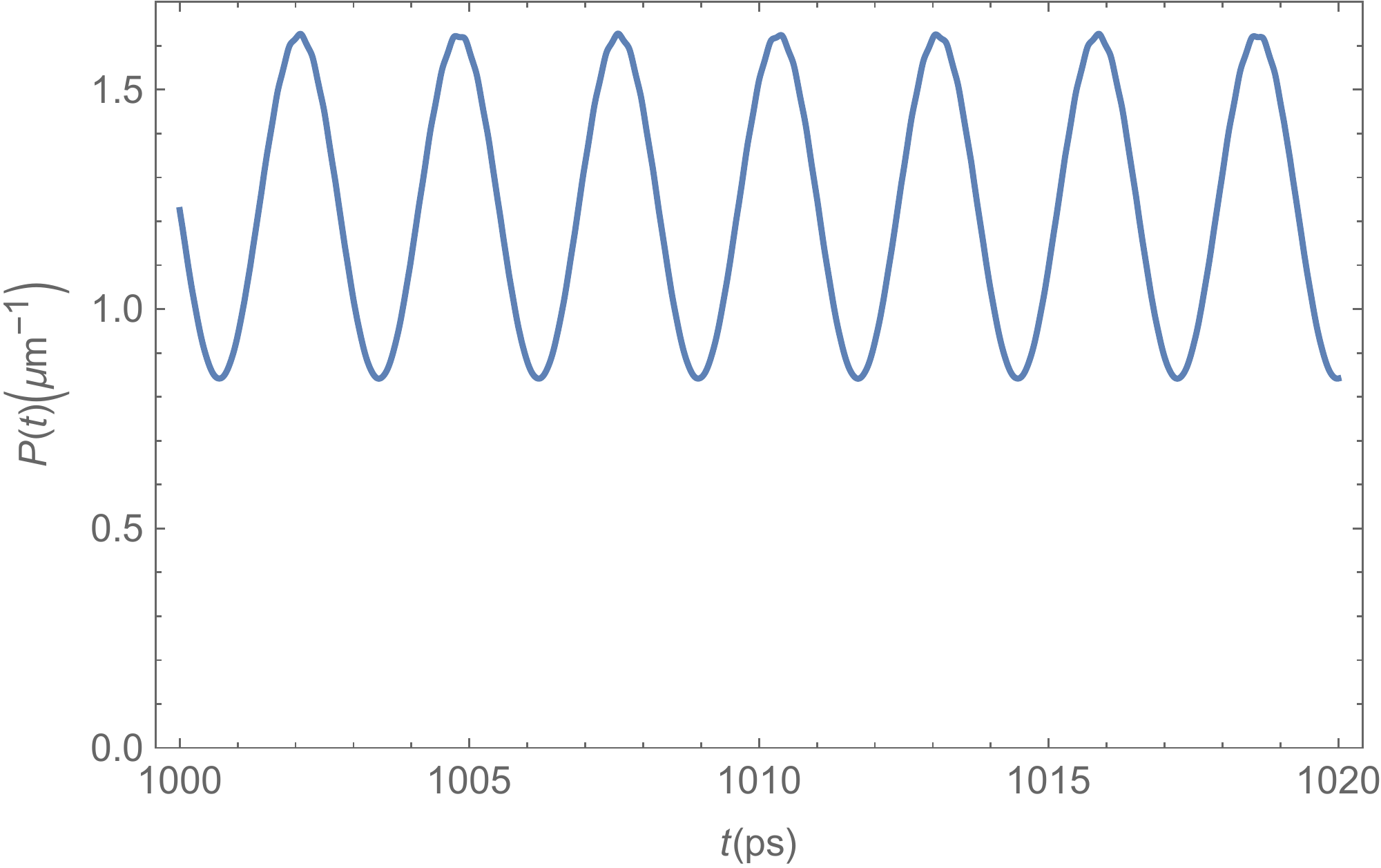}
\caption{}
\end{subfigure}
\hfill
\begin{subfigure}[b]{0.32\textwidth}
\includegraphics[width=\textwidth]{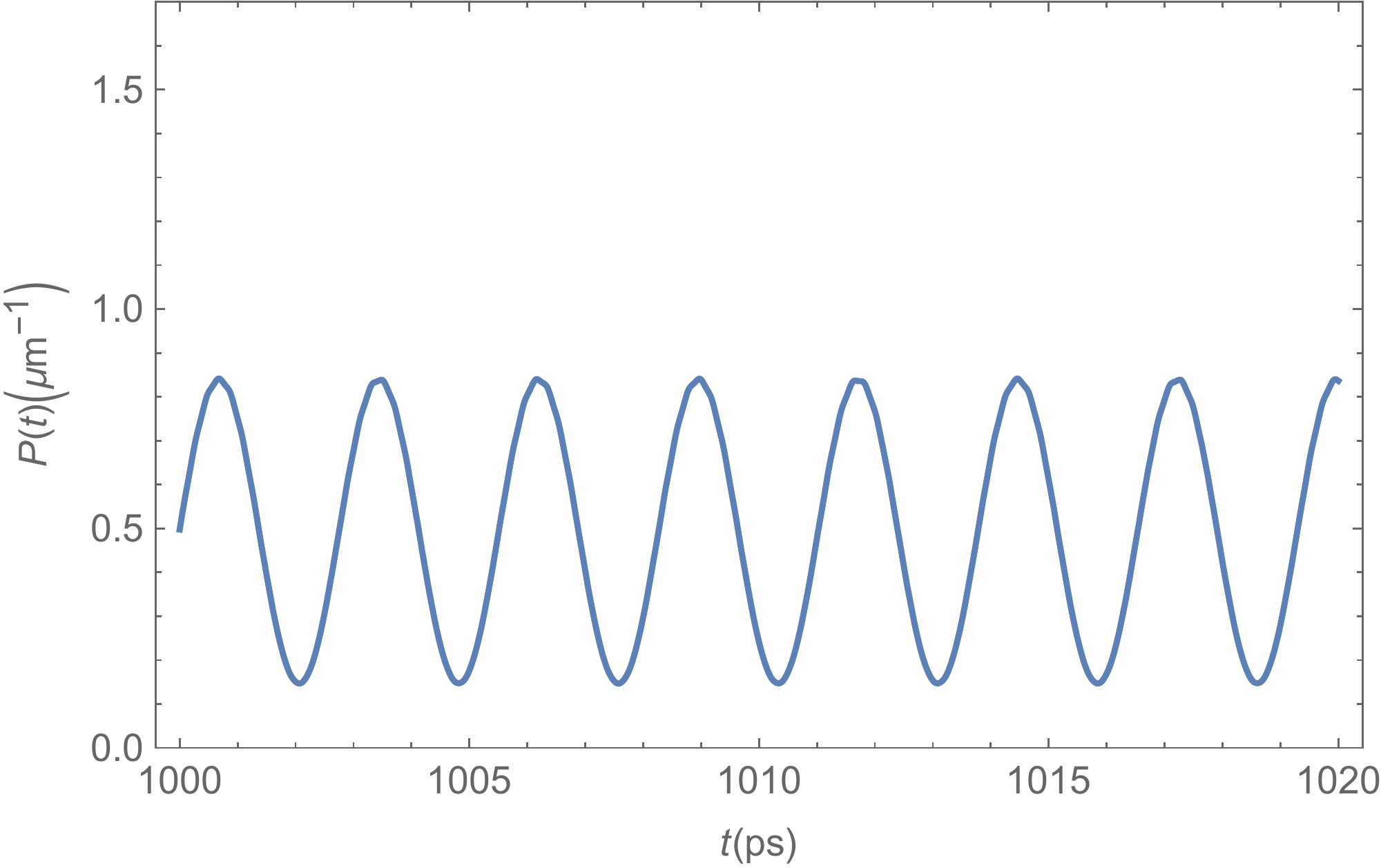}
\caption{}
\end{subfigure}
\hfill
\end{center}
\caption{Polariton BEC density $P(t,x) = |\varphi(t,x)|^2 $ {in} a ribbon of length 4000 $\mu$m at {three} chosen positions. (a)  x = $\pm$ 125.0  $\mu$m;  (b) x = $\pm$ 250.0 $\mu$m; and (c)  x = $\pm$ 375.0 $\mu$m.}
\label{positions}
\end{figure}

{In Fig. \ref{positions}, we see the time evolution of the polariton BEC density in three different positions in the strip. The condensate density oscillates between $P_0$ and a value that can be smaller than $P_0$, as in Figs. \ref{positions} (a) and \ref{positions} (c), or greater than $P_0$, as in Fig \ref{positions} (b). The system oscillates in phase, reaching $P_0$ at the same time for all positions; and reaching the maximum deviation also at the same time. The period of oscillation can be seen to be of around 3 ps. We, again, see that the system does not show any tendency of reaching a steady state. Those results are completely compatible to those shown in Fig. \ref{times}, as it should be expected.  }

{In order to check if our result {depends} on the initial conditions chosen by us, namely the steady-state for the unperturbed BEC (when $V_{\mathrm{eff}}=0$), we tried a different initial condition, when the system started in the vacuum state. This result is shown in Fig. \ref{vac}.}

{ On the top row of Fig. \ref{vac}, we see a picture of the condensate density throughout the entire strip at an arbitrary time, just like one of the panels in Fig. \ref{times}. On the bottom row, we see the time evolution of the condensate density at an arbitrary position, just like in of the panels in Fig. \ref{positions}. In Fig \ref{vac} (a), the system started at $t=0$ in the steady-state of the unperturbed system, like on all previous and future Figures, while in Fig. \ref{vac} (b), the system started in the vacuum state. It is evident that the only difference between the figures is the overall phase of the system. The dynamics themselves, for $t$ long enough, are completely equivalent. This means that our results were not impacted by our choice of initial state and should be verifiable regardless of how the system is at $t=0$. }

{ So far, we have shown evidence that our system \textit{could} be in a TC phase, but haven't yet provided the definite proof, through the mathematical criterion proposed by Watanabe and Oshikawa \cite{Watanabe}. This will be done now. }
We will show that our system obeys Eq.~(\ref{cond}). In contrast, many of the proposed TCs in the literature  have not shown that they satisfy such a condition \cite{Savona,Kavokin}. 

We consider the observable $\hat{O}(x,t)$ to be the deviation {of the BEC density} from  the unperturbed steady-state density, $\hat{O} = P (x,t)-P_0$.

\begin{figure}[H]
\begin{center}
\includegraphics[width=0.5\textwidth]{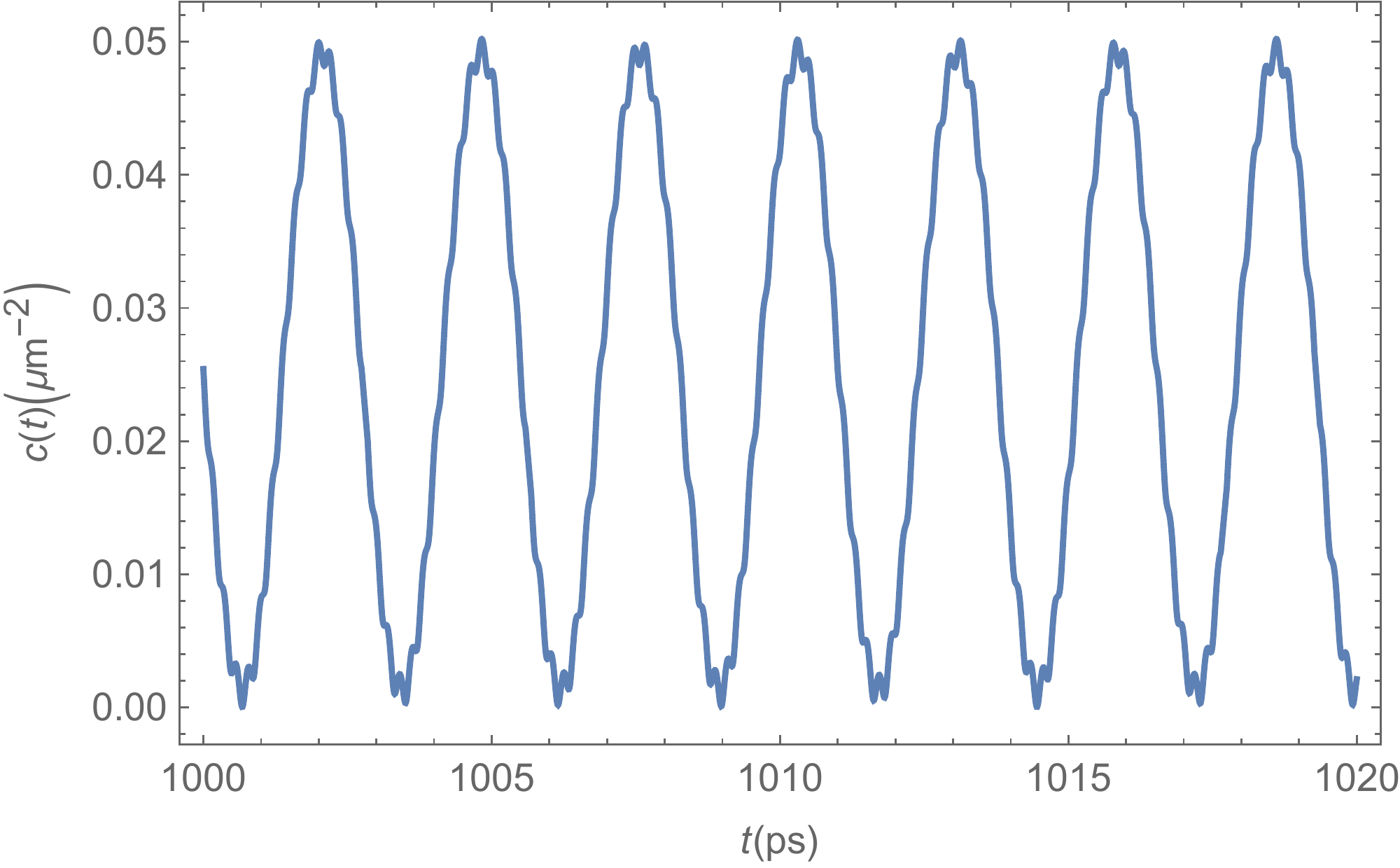}
\end{center}

\caption{{Two-point correlation function $c(t)$ as defined in Eq. \ref{cond} {for the polariton BEC.} For this plot, we considered the observable $O$ to be the deviation between the condensate density $P(x,t)$ and the unperturbed steady-state density {$P_0$}. We took $x-x^\prime$ to be half the length of the strip and calculated the expectation value $\left\langle \hat{O}(\mathbf{r},t) \hat{O}(\mathbf{r}^\prime,t^\prime) \right\rangle$ by taking the average value for all $x$ while maintaining $x-x^\prime$ constant. \label{son} }}
\end{figure}

It can be seen from Fig. \ref{son} that {the polariton BEC in a strip of TMDC, embedded in a microcavity  in the presence of the external periodic potential}, in fact, obeys the mathematical condition given by Eq. (\ref{cond}) {within the mean-field approach}, and can be, therefore, characterized as a Time Crystal. A brief comparison between the two-point correlator shown in Fig \ref{son} with the time evolution of the polariton condensate shown in Fig. \ref{positions} shows us that this correlator oscillates in time with the same frequency as the condensate itself oscillates, a result that does seem reasonable.

\subsubsection{Beyond Mean-Field description}

{ We now turn our attention to the BEC dynamics when quantum corrections are taken into consideration. The results we show here depict {the numerical solution of the} time evolution of the polariton BEC density following the DSGPE in Eq. (\ref{DSGPE}). }

{

We expect Eq.\  (\ref{feex}) to hold as long as the stochastic term obeys $ \left(\kappa + \gamma \right) \psi_{Q}\ll V_0$, $\dfrac{U}{2} \left|\psi_0\right|^{2}$, $\frac{\hbar^2}{2M}\left|\psi_0\right|^{2}$, where $\left|\psi_0\right|^{2} = \dfrac{\kappa}{\Gamma}$ is the unpertured condensate density of the steady state. In our considerations, these values vary from 0.65 meV to 3.1 meV. Therefore, the  correction is  a whole order of magnitude smaller   than the other terms and may be treated as a perturbation. In Fig.\  \ref{beyond}, we see the time evolution of the condensate density in various positions along the  ribbon, {similarly to Fig. \ref{positions}, for the mean-field approach.}

\begin{figure}[H]
\begin{center}
\hfill
\begin{subfigure}[b]{0.32\textwidth}
\includegraphics[width=\textwidth]{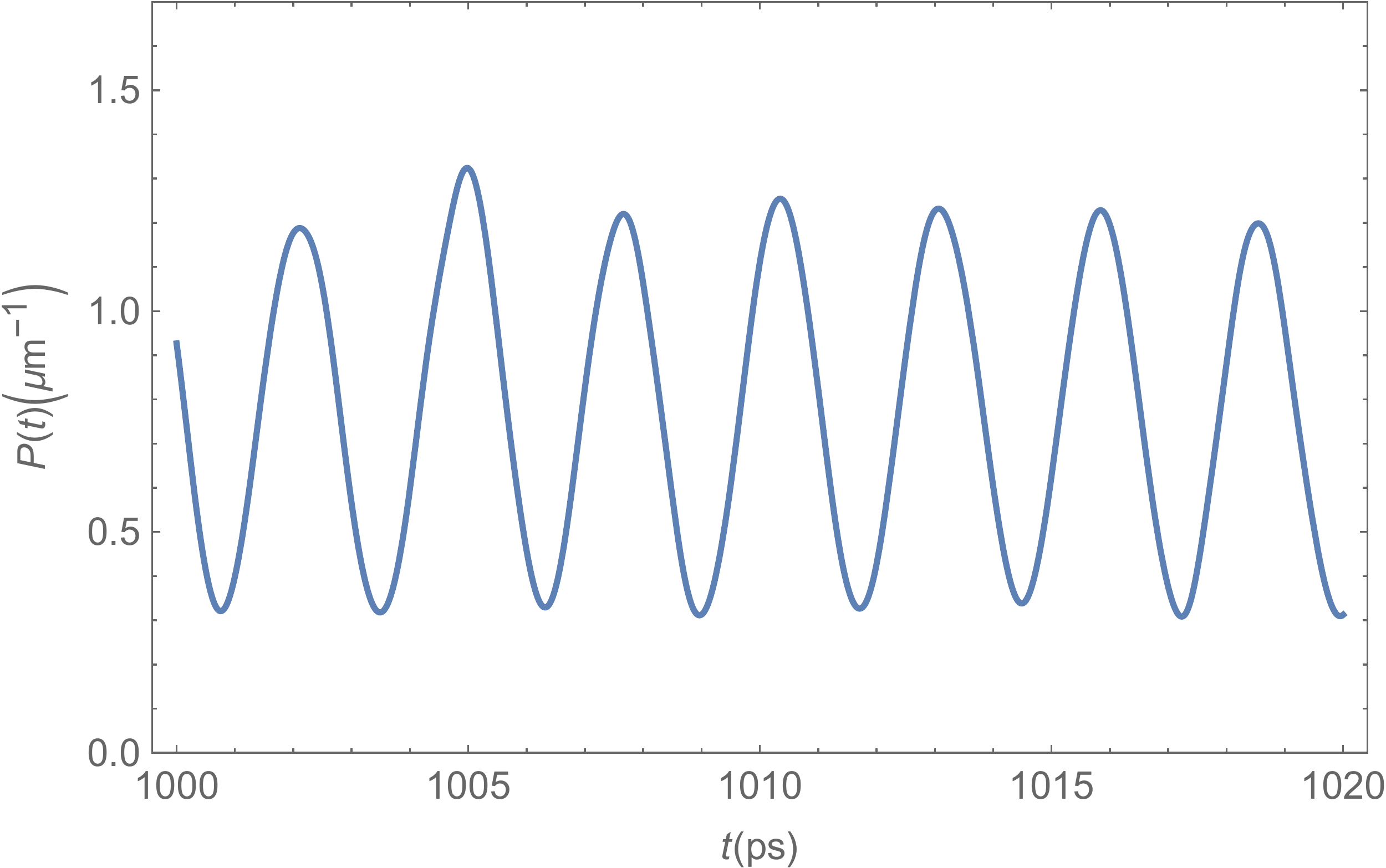}
\caption{}
\end{subfigure}
\hfill
\begin{subfigure}[b]{0.32\textwidth}
\includegraphics[width=\textwidth]{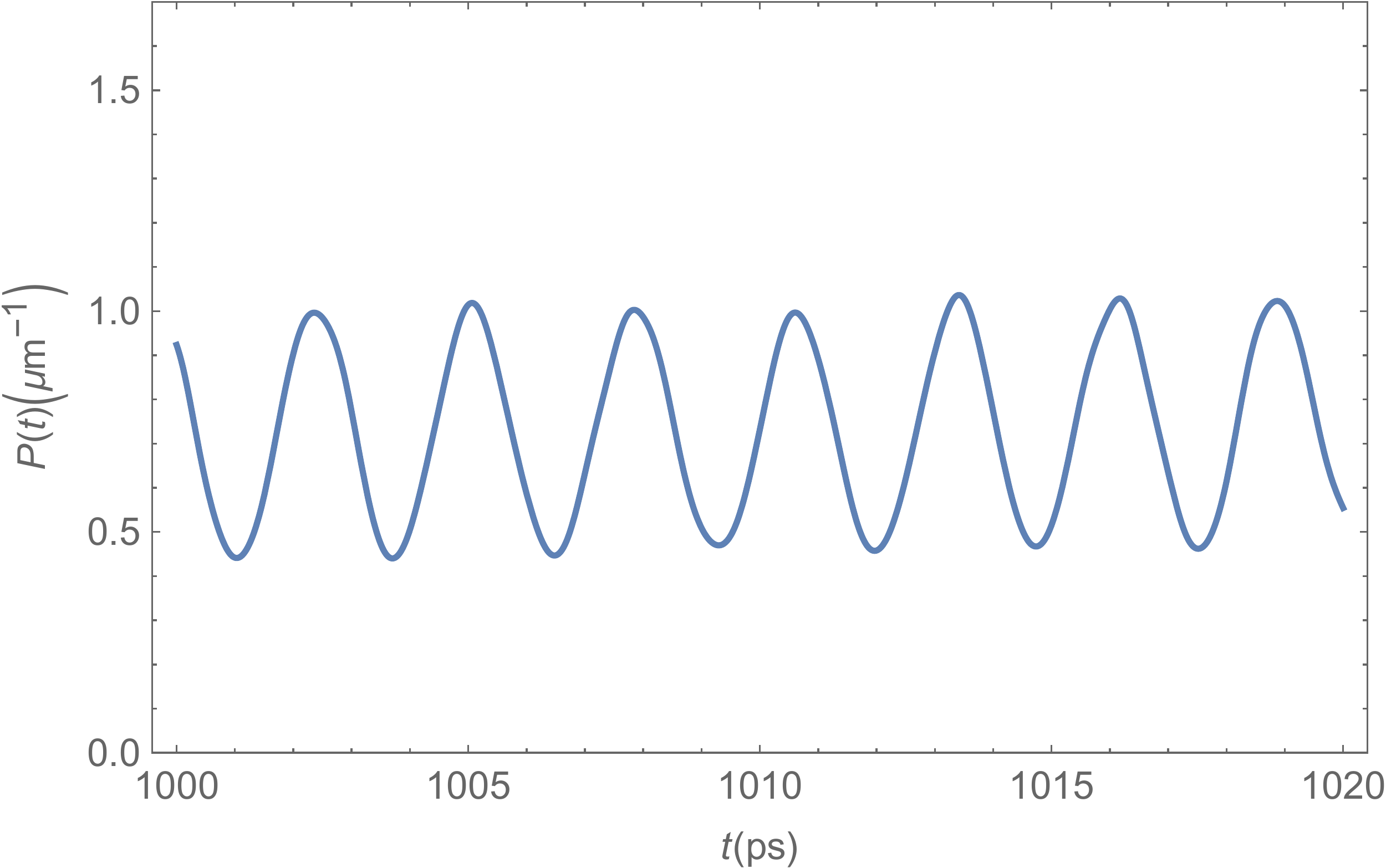}
\caption{}
\end{subfigure}
\hfill
\begin{subfigure}[b]{0.32\textwidth}
\includegraphics[width=\textwidth]{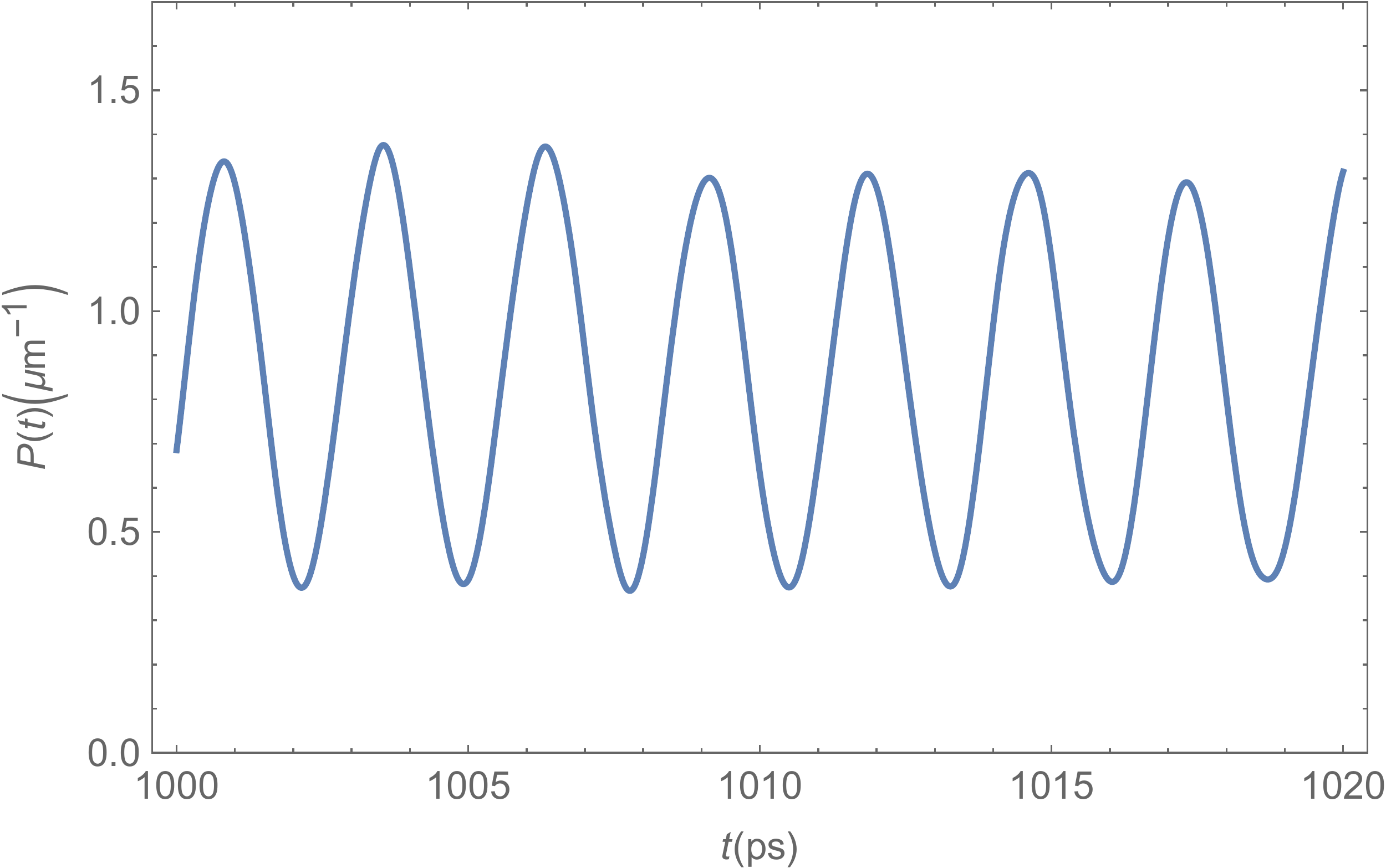}
\caption{}
\end{subfigure}
\end{center}
 \caption{Polariton {BEC} density $P(t,x) = |\varphi(x,t)|^2 $  {in} a strip {4000} $\mu$m long at {three} different positions as a function of time, {when quantum uncertainty is taken into consideration.} (a) x = 125 $\mu$m; (b) x = 250 $\mu$m; and (c) x = 375 $\mu$m.}\label{beyond}
\end{figure}

As one can see in Fig.\   \ref{beyond}, the addition of noise due to quantum uncertainty in the calculation does, in fact, break the perfect periodicity seen in the mean-field analysis. {Unlike in Fig. \ref{positions}, the actual maximum and minimum values for the BEC density at each oscillation are not exactly the same at each oscillation, but change slightly. Other than that, the exact times at which each position reaches a peak and a valley in the condensate density also varies by a tiny amount.} However, in this regime where pumping and decay are not sufficiently strong (low density regime), this ``quantum noise" is not strong enough  to erase the overall shape of the mean-field evolution and we still see clear evidence of a Time Crystal phase in the system. In order to determine once and for all that the system is, in fact, still a TC, one needs only to check if it still obeys the mathematical condition given by Eq. (\ref{cond}). {This result will be shown in Fig. \ref{son2}.}

\begin{figure} [H]
\begin{center}
\includegraphics[width=0.5\textwidth]{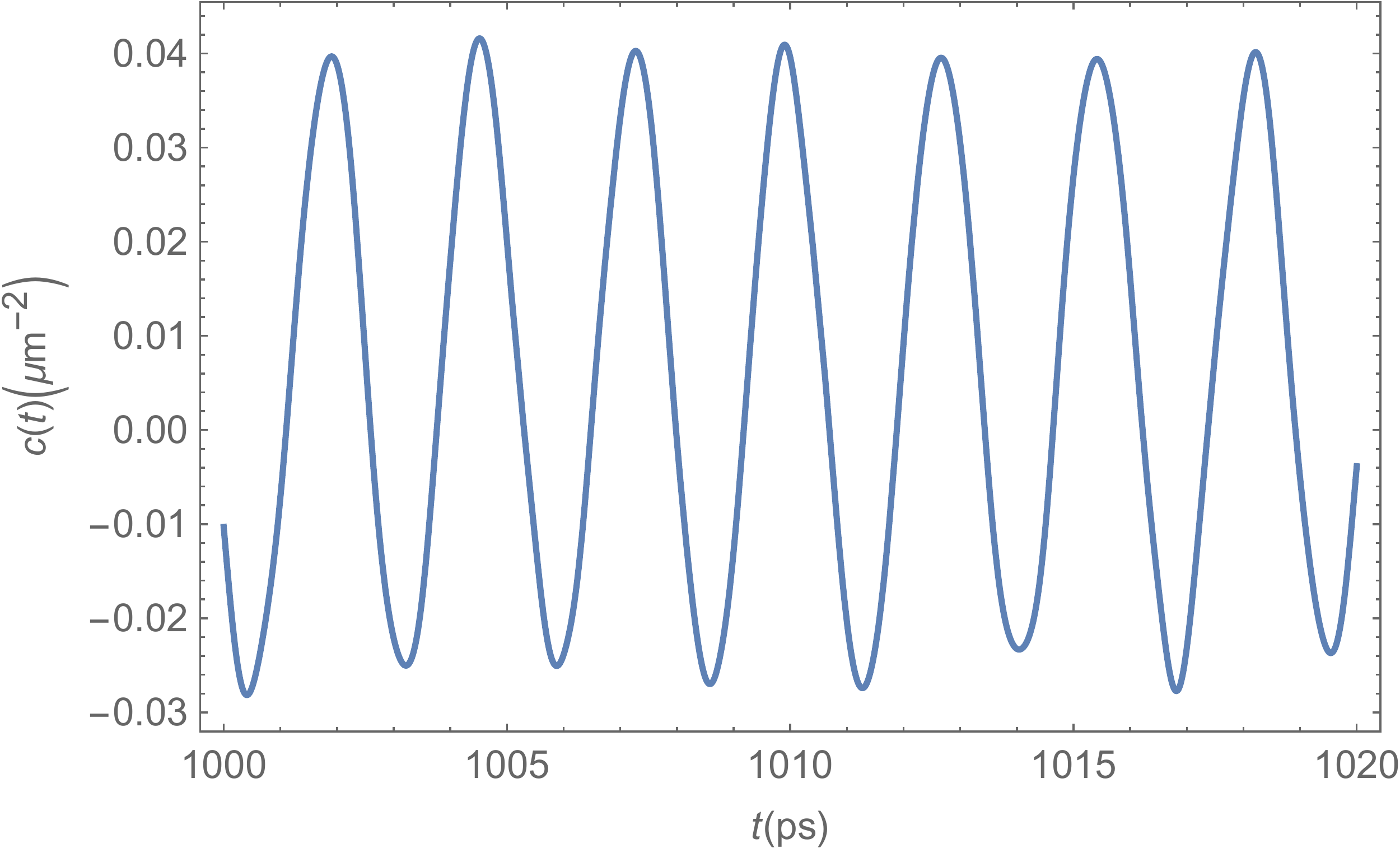}
\end{center}

\caption{{Two-point correlation function $c(t)$ as defined by Eq. (\ref{cond}) {for the exciton-polariton condensate, with the addition of} quantum noise. { Like in Fig. \ref{son}, we considered the observable $O$ to be the deviation between the condensate density $P(x,t)$ and the unperturbed steady-state density {$P_0$}. We took $x-x^\prime$ to be half the length of the strip and calculated the expectation value $\left\langle \hat{O}(\mathbf{r},t) \hat{O}(\mathbf{r}^\prime,t^\prime) \right\rangle$ by taking the average value for all $x$ while maintaining $x-x^\prime$ constant.} \label{son2} }}
\end{figure}

By looking at Fig. \ref{son2}, which shows us the two-point correlation function of the deviation between the polariton density and the unperturbed steady-state density at different positions and times, it is evident that this correlator obeys the mathematical condition presented in Eq. (\ref{cond}), which proves definitively that the system is in a time crystaline phase.  A comparison between Figs. \ref{son} and \ref{son2} leads to an interesting conclusion: The two-point correlator that proves the system is, in fact, in a TC phase is actually enhanced on the case with quantum noise. With the deterministic GPE, this correlator is always positive and oscillates between zero and 0.05 $\mu$m$^{-2}$. For the DSGPE which takes into consideration quantum uncertainties, in the other hand, we see this correlator becoming negative at some points and presenting oscillation that have about twice the amplitude, between -0.04 and 0.07  $\mu$m$^{-2}$. The overall width of the peaks is, however, unaltered. Since this correlator serves as proof of the TC phase, one could argue that the TC phase does not simply perseveres with the addition of quantum noise, but rather becomes actually more evident.  { Therefore, the polariton BEC in a strip of TMDC, embedded in a microcavity  in the presence of the external periodic potential is a Time Crystal under consideration beyond the mean field approach.}

}

\begin{figure}[H]
\begin{center}
\begin{subfigure}[b]{0.30\textwidth}
\includegraphics[width=\textwidth]{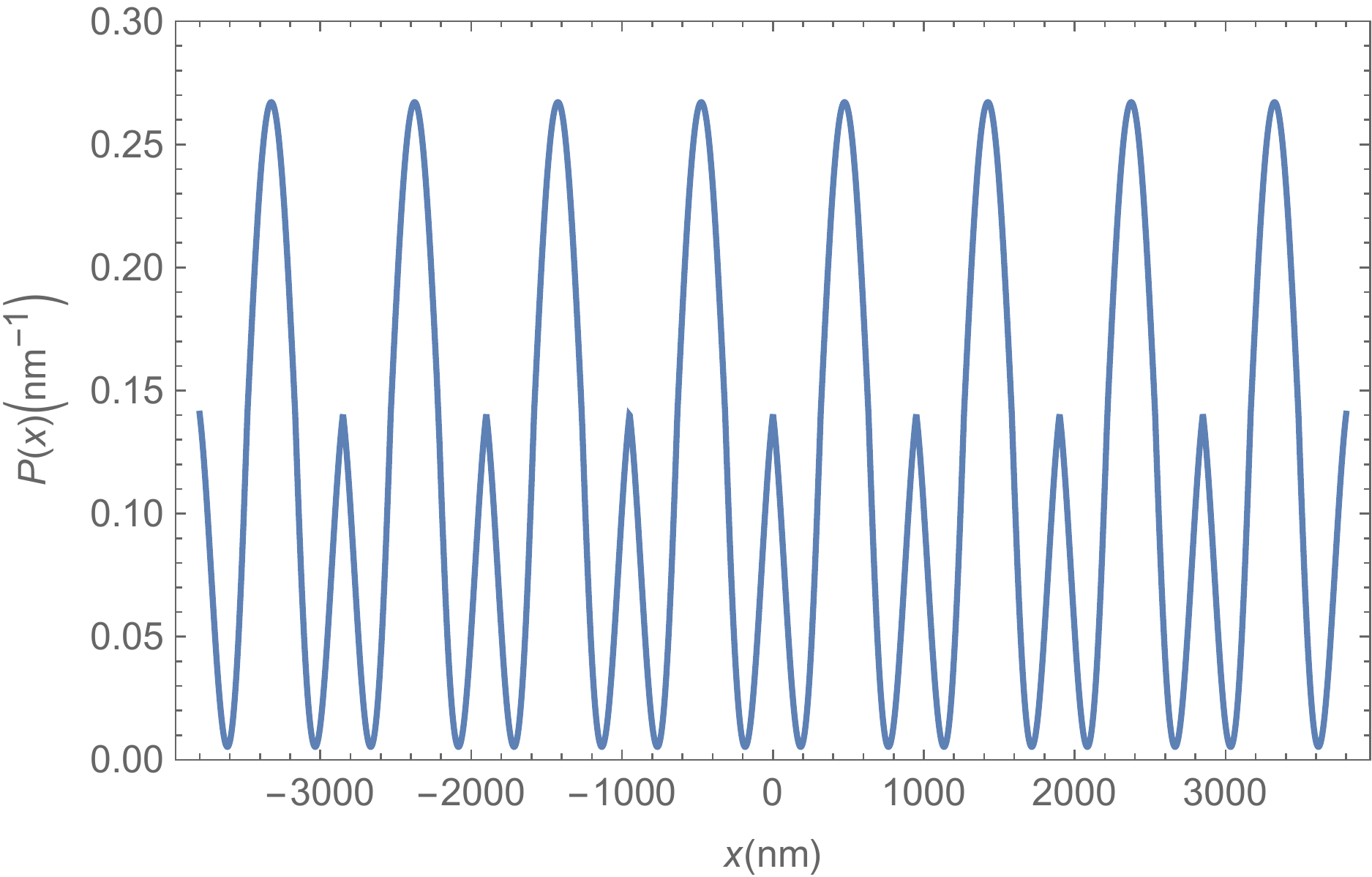}
\caption{}
\end{subfigure}
\hfill
\begin{subfigure}[b]{0.30\textwidth}
\includegraphics[width=\textwidth]{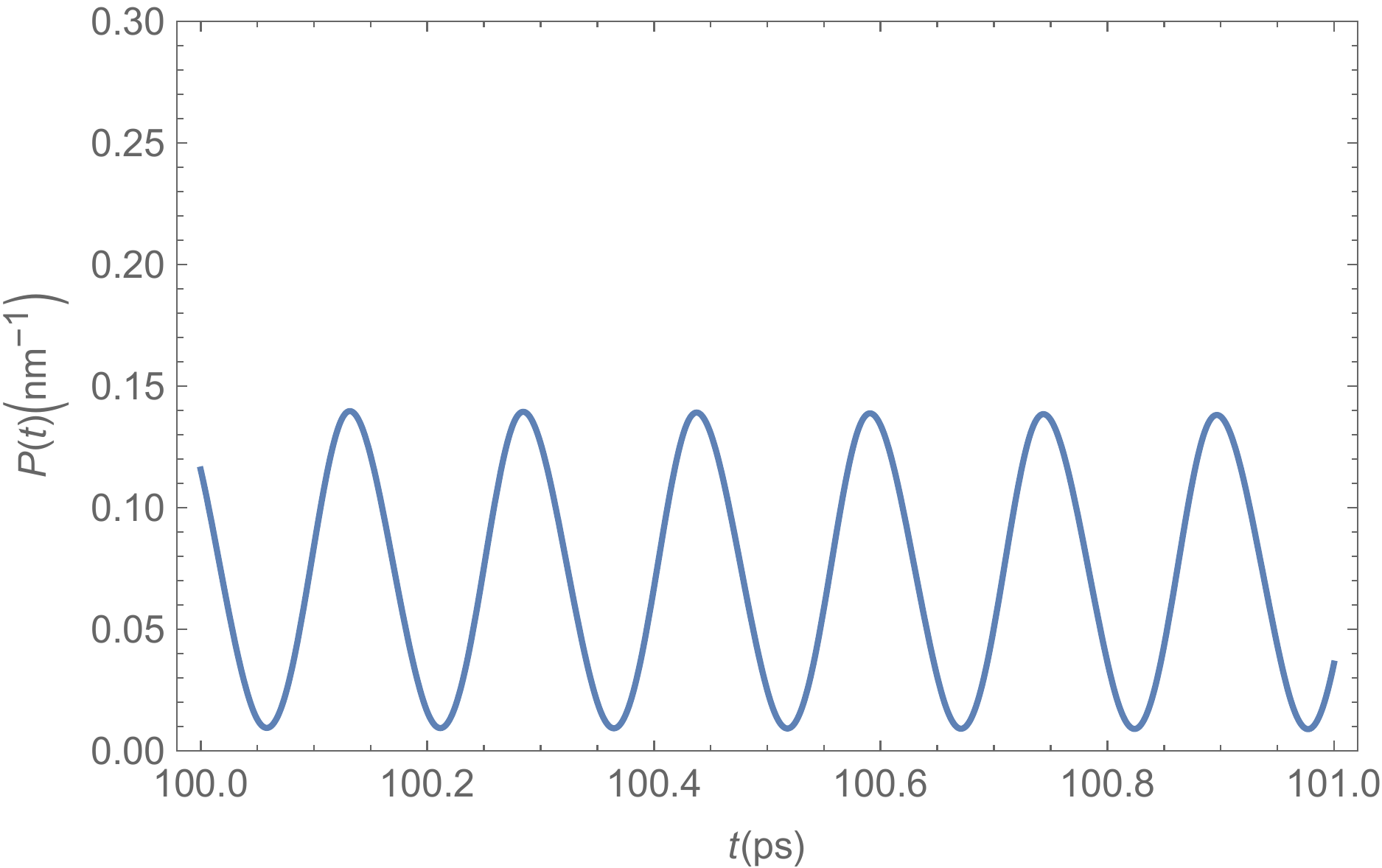}
\caption{}
\end{subfigure}
\hfill
\begin{subfigure}[b]{0.30\textwidth}
\includegraphics[width=\textwidth]{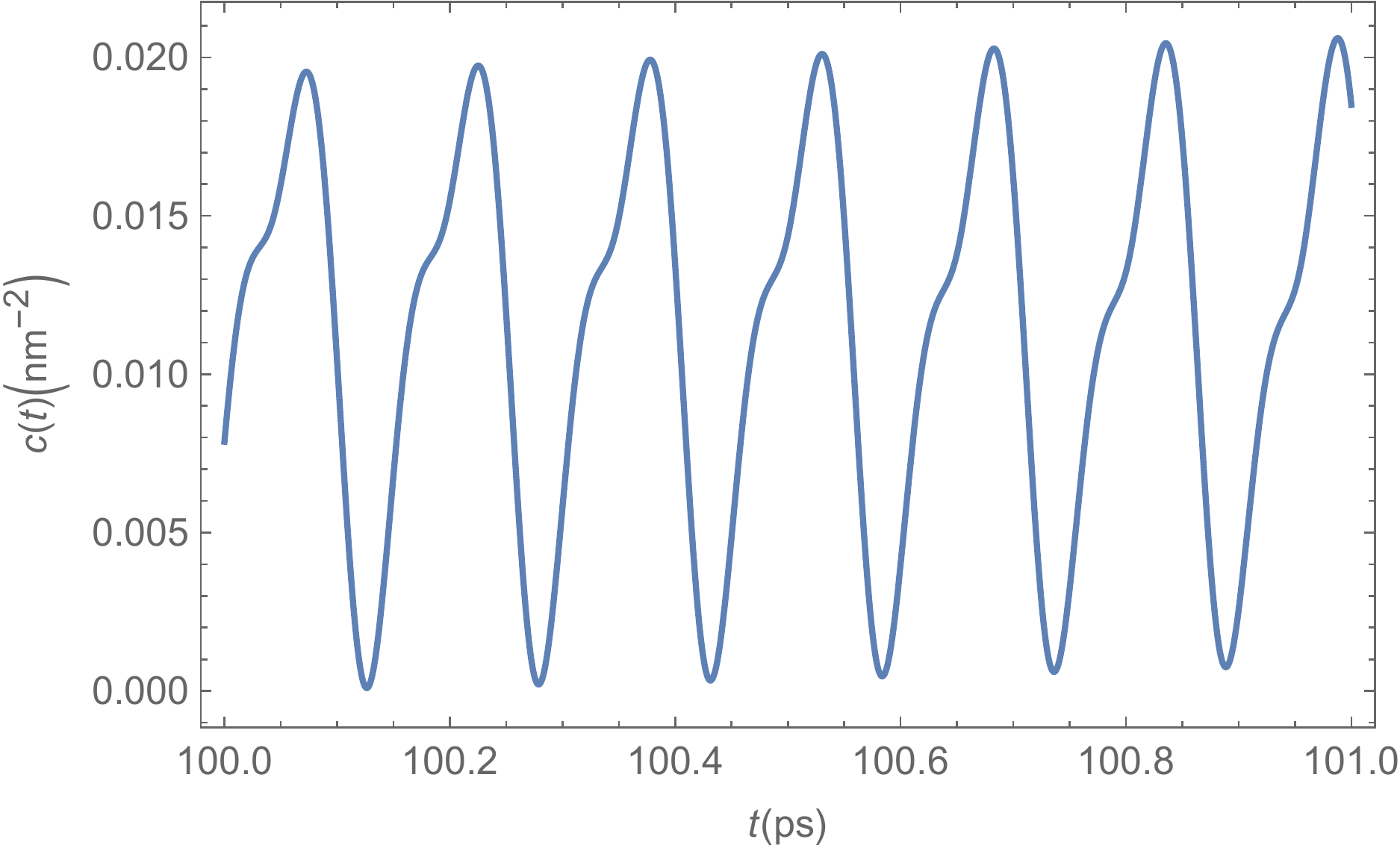}
\caption{}
\end{subfigure}
\end{center}

\caption{{ Numerical results for the dynamics of the excitonic condensate{in a twisted TMDC bilayer in the mean-field approach}. (a) time evolution of the condensate density at position $x= 100$ nm; (b) Condensate density throughout the entire strip at $t=3$ ns and (c) two-point correlator as defined in Eq. (\ref{cond}). \label{extc}}  }

\end{figure}

\begin{figure}[H]
\begin{center}
\begin{subfigure}[b]{0.30\textwidth}
\includegraphics[width=\textwidth]{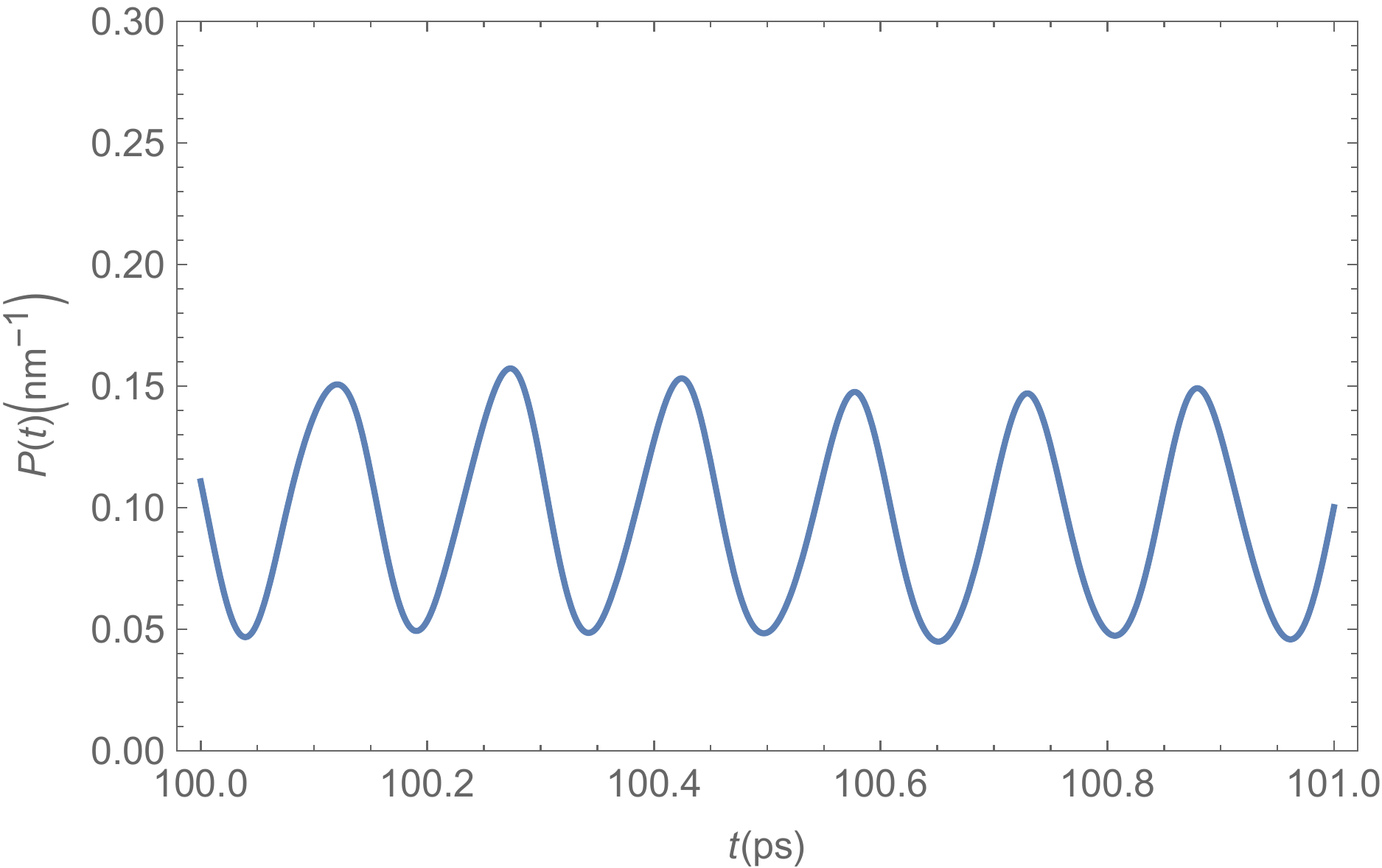}
\caption{}
\end{subfigure}
\begin{subfigure}[b]{0.30\textwidth}
\includegraphics[width=\textwidth]{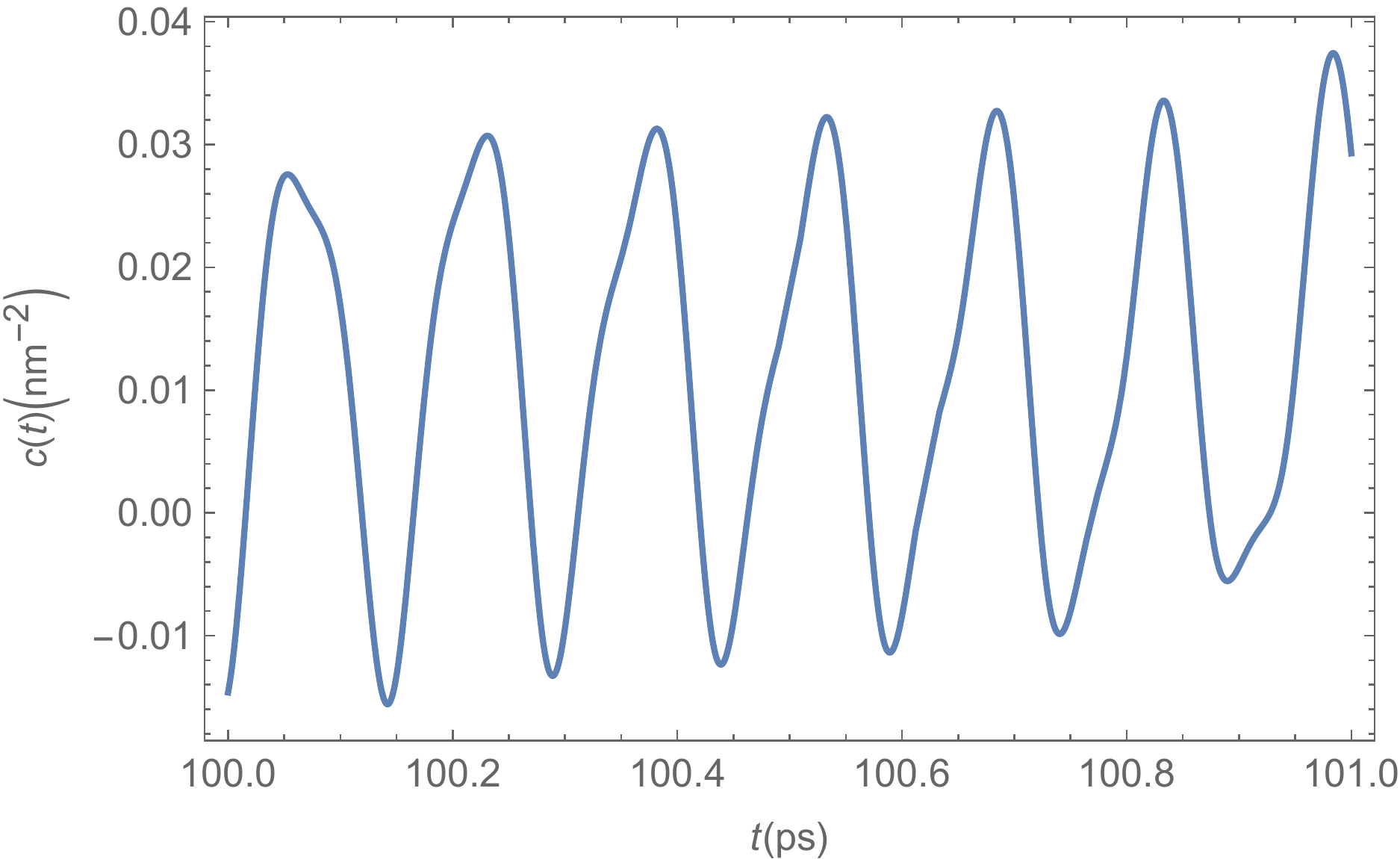}
\caption{}
\end{subfigure}
\hfill
\end{center}
\caption{{Numerical results for the dynamics of the excitonic condensate {in a twisted TMDC bilayer} beyond the mean-field approach. (a) Condensate density in an arbitrary position of the strip. (b) Two-point correlator defined by Eq. (\ref{cond}) }\label{extb}}
\end{figure}

{
\subsection{Excitonic BEC on a twisted TMDC bilayer}

We will now present our results for the {mean-field dynamics of the} bare exciton BEC {in a strip of twisted TMDC bilayer}. In order to avoid seemingly repetitive plots and discussions, we will compile all results for this condensate in { two figures. In Fig. \ref{extc}, we combine all the results for the mean-field dynamics of the condensate. In Fig. \ref{extb}, we present our results for the dynamics of the condensate when quantum uncertainty is taken into consideration.}

}

We considered the same system as in \cite{Tran}, namely a WSe$_2$/MoSe$_2$ bilayer heterostructure twisted by 1$^\circ$, we will have $M_{ex}\approx 0.84$ $m_0$, {$a_M\approx19$ nm} and $V_0\approx 18$ meV. { We considered our strip to be 8 $\mu$m long and to have a width of $10.4$ nm, which obeys the criterion for the effective potential to be given by Eq. (\ref{vexc}).} The resulting dynamics of the excitonic condensate and two-point correlator can all be seen in Fig. \ref{extc}. 

{ Each of the panels in Fig. \ref{extc} {shows results for the mean-field evolution of the excitonic BEC in a twisted TMDC bilayer that are} { similar} to one of the figures in the mean-field analysis of the polariton BEC. Fig. \ref{extc} (a) shows the BEC {mean-field} density throughout the entire strip, just like Fig. \ref{times} did for the polaritons. The similarities are evident, the overall shape of the BEC is completely equivalent, the only difference being the scale. This comes from the fact that the bare exciton mass is about six orders of magnitude larger than the effective mass of the polaritons, which leads to the bare exciton BEC {length} scales to be about three orders of magnitude smaller than those of the polariton BEC. Figure \ref{extc} (b) shows the time evolution of the condensate {mean-field} density in an arbitrary position as a function of time, just like what we saw in the panels of Fig. \ref{positions} for the polariton BEC. Those results are completely equivalent, just with a different period of oscillation, which is about one order of magnitude smaller. Lastly, in Fig. \ref{extc} (c), we show the two-point correlator from Eq. (\ref{cond}). It is evident that this correlator also obeys Watanabe and Oshikawa's criterion{, given in Eq. (\ref{cond}),}  and {the exciton BEC in a twisted TMDC bilayer is}, therefore, a TC.}

{In Fig. \ref{extb}, we see numerical results for the time-evolution of the condensate density in an arbitrary position (panel (a)); and the two-point correlator defined in Eq. (\ref{cond}) (panel (b)). Just like in the polaritonic condensate beyond the mean-field description, we lose the perfect definition of the oscillations when quantum uncertainty is taken into consideration. The actual height of each peak and valley in Fig \ref{extb} (a) is slightly different from one another. This, however, is not nearly enough to erase the overall shape seen by the mean-field dynamics in Fig. \ref{extc} (b). The comparison between the two-point correlator beyond mean-field shown in Fig. \ref{extb} (b) with its mean-field counterpart, shown in Fig \ref{extc} (c) is also very similar to the polaritonic condensate case. The overall shape of this correlator does change visibly, but its shape still shows clear signs of oscillations in time. The intensity of the peaks was greatly enhanced in the non-deterministic beyond mean-field approach, when compared to the deterministic mean-field approach and, like in the polaritonic BEC, we now see negative values for the correlator that were not observed in the mean-field approach. However, the criterion proposed in Eq.  (\ref{cond}) is still obeyed and the system is, therefore, still in a TC phase. As we can see from the panels of Fig. \ref{extb}, the results obtained for the mean-field dynamics of the excitonic BEC still hold even when quantum uncertainty is taken into consideration, just like for the polaritonic BEC.}

\section{Concluding Remarks}
\label{conc}

{Throughout this paper, we have studied in depth the dynamics of   BECs of exciton-polaritons and of bare excitons on a strip of TMDC {in the presence of} an external periodic potential. {We started by presenting the theoretical foundation of the paper. We did this by thoroughly describing a system of exciton-polaritons in a strip of TMDC inside a spatially curved optical microcavity. We explained how the curvature of the microcavity leads to an effective periodic potential for photons inside the cavity, which translates into an effective potential for the polaritons themselves. We also did a similar explanation of a system of bare excitons on a strip of twisted TMDC bilayer, in which a similar effective potential to that of the curved microcavity arose from the Moiré pattern created by the twist.  After that, we presented the mathematical criterion that should be used to verify whether a system is in a Time Crystalline phase. After this thorough explanation of the theoretical framework we were building upon, we defined the Gross-Pitaevskii equation that governs the dynamics of the BECs mean-field density and showed how to turn the originally 2D GPE into the simpler one-dimensional form, needed for a narrow strip. We also showed how to add stochastic corrections to the deterministic GPE consistent with the intrinsic quantum randomness that is present in the dynamics, going beyond the usual mean-field description.} Lastly, we presented several results for the time-evolution of {the polariton BEC density, showing that the condensate throughout the strip oscillates in time.  In summary, the condensate density oscillates in phase throughout the strip, even though the maximum and minimum values of the condensate density vary from point-to-point. After that, we calculated the two-point correlator that Watanabe and Oshikawa \cite{Watanabe} proposed to serve as a criterion to verify whether a system is in a TC phase or not, and proved our system obeys it which is something that was not done in other propositions of TCs based on BECs \cite{Kavokin,liao,wright,wang}. We then presented simulations beyond the usual mean-field description of the BEC, and showed that our results still hold, even when we take into consideration quantum uncertainty.  Next, we presented our results for the BEC of bare excitons on a strip of twisted TMDC, which were, as expected, {very similar} to those of exciton-polaritons on a spatially curved microcavity, proving that this system also behaves as a TC.

\medskip
\par
{ It is our opinion that both of our considered systems are good candidates for future experimental verification. Since recently a BEC of exciton-polaritons has been verified in room-temperature settings in halide perovskite \cite{pol_fluid}, it is reasonable to assume the same might be possible for TMDCs. If that is the case, our results lead to the possibility of having a room-temperature TC, which would depend only on the manufacturing of our proposed  spatially curved microcavity. Our second system, namely bare-excitons on a twisted TMDC bilayer, on the other hand, has been seen to form condensates at temperatures around 190 K \cite{excond}, way above the ultra-cold regime in which atoms form condensates and reachable in most research settings. Such a system has the advantage that the external potential arises naturally from the twisting of the TMDC {layers, forming the} bilayer \cite{Tran}, making the manufacture process much simpler.}
}}



\medskip

\par

\section*{Acknowledgement(s)}

{The authors are grateful to  V.~S. Boyko and O. V. Roslyak}  for the valuable discussions. The authors are grateful for support by grants: G.P.M. and O.L.B. from U.S. ARO grant No. W911NF1810433. GG acknowledges the support from the US AFRL Grant No. FA9453-21-1-0046.


\begin{thebibliography}{99}


\bibitem{Wilczek} F. Wilckzek, Quantum Time Crystals, \prl {\textbf{ 109}}, 160401 (2012).


\bibitem{Watanabe} H. Watanabe and M. Oshikawa, Absence of Quantum Time Crystals, \prl  {\textbf{ 114}}, 251603 (2015).

\bibitem{Noz} P. Nozières, Time Crystals: Can diamagnetic currents drive a charge density wave into rotation?, Europhys. Lett. {\textbf{ 103}}, 57008 (2013).

\bibitem{Bruno} P. Bruno, Impossibility of spontaneously rotating Time Crystals: a no-go theorem \prl  {\textbf{ 111}}, 070402 (2013).


\bibitem{Khemani} V. Khemani, A. Lazarides, R. Moessner  and S. L. Sondhi, Phase structure of driven quantum systems, \prl \textbf{ 116}, 250401 (2016).

\bibitem{Else} D. V. Else, B. Bauer and C. Nayak, Floquet Time Crystals, \prl \textbf{ 116}, 250401 (2016).

\bibitem{Keyser} C. W. von Keyserlingk, V. Khemani and S. L. Sondhi, Absolute stability and spatiotemporal long-range order in Floquet systems, \prb \textbf{ 94}, 085112 (2016).

\bibitem{Sondhi} V. Khemani, R. Moessner, S. L. Sondhi, A brief history of Time Crystals, \textit{preprint}, arXiv:1910.10745 (2019).

\bibitem{spin} M. Medenjak, B. Buča, and D. Jaksch, Isolated Heisemberg magnet as a quantum Time Crystal \prb \textbf{ 102}, 041117(R) (2020).



\bibitem{Kavokin} A.~V. Nalitov, H. Sigurdsson, S. Morina, Y.~S. Krivosenko, I.~V. Iorsh, Y.~G. Rubo,
A.~V. Kavokin, and I.~A. Shelykh, Optically trapped polariton condensate as a semiclassical time crystal \pra   \textbf{ 99}, 033830 (2019).




\bibitem{liao} L. Liao, J. Smits, P. van der Straten, and H. T. C. Stoof, Dynamics of a dpace-time crystal in an atomic Bose-Einstein condensate \pra \textbf{ 99}, 013625 (2019).

\bibitem{wright} P. Öhberg and E. M. Wright, Quantum time crystals and interacting gauge theories in atomic Bose-Einstein condensates, \prl \textbf{123}, 250402 (2020).

\bibitem{wang} J. Wang, P. Hannaford, and B. J. Dalton, Many-body effects and quantum fluctuations for discrete time crystals in Bose–Einstein condensates, New J. Phys. \textbf{ 23} 063012 (2021).


\bibitem{kozin} V. K. Kozin and O. Kyriienko, Quantum time crystals from Hamiltonians with long-range interactions, \prl {\textbf{ 123}}, 210602(2019).

\bibitem{Savona} K. Seibold, R. Rota, and V. Savona, Dissipative time crystal in an asymmetric nonlinear photonic dimer,  \pra {\textbf{ 101}}, 033839 (2020).


\bibitem{brain} P. Singh, K. Saxena, A. Singhania, P. Sahoo, S. Ghosh, R. Chhajed, K. Ray, and D. Fujita, A Self-Operating Time Crystal Model of the Human
Brain: Can We Replace Entire Brain Hardware with a
3D Fractal Architecture of Clocks Alone?,  Information \textbf{11} (5): 238 (2020).



\bibitem{Moskalenko_Snoke} S.~A. Moskalenko and D.~W. Snoke, \textit{%
Bose-Einstein Condensation of Excitons and Biexcitons and Coherent
Nonlinear Optics with Excitons} (Cambridge University Press, New
York, 2000).


\bibitem{excond} A. Kogar \textit{et al.} Signatures of exciton condensation in a transition metal dichalcogenide, Science \textbf{ 358}, 6368 1314-1317 (2017).


\bibitem{Snoke_Keeling} D.~W. Snoke and J. Keeling, The new era of polariton condensates,  Physics Today {\bf 70}, 54 (2017).


\bibitem{Littlewood} P. Littlewood, Condensates made of light, Science \textbf{316}, 989 (2007).


\bibitem{Carusotto_rmp} I. Carusotto and C. Ciuti, Quantum fluids of light, \rmp \textbf{ 85}, 299 (2013).



\bibitem{Snoke_book} N. Proukakis, D.~W. Snoke, and P.~B. Littlewood, \textit{Universal Themes of Bose-Einstein Condensation}%
, (Cambridge University Press, 2017).


\bibitem{pol_fluid} K. Peng, R. Tao, L. Haeberlé,  \textit{et al.} Room-temperature polariton quantum fluids in halide perovskites, Nat. Commun. \textbf{13}, 7388 (2022).




\bibitem{Kormanyos} A. Korm\'{a}nyos, G. Burkard, M. Gmitra, J. Fabian, V. Z%
\'{o}lyomi, N.~D. Drummond, and V. Fal'ko, k$\cdot$p theory for two-dimensional transition metal dichalcogenide semiconductors, 2D Mater. \textbf{2},
022001 (2015).

\bibitem{Mak2010} K.~F. Mak, C. Lee, J. Hone, J. Shan, and T.~F. Heinz, Atomically Thin MoS$_2$: A New Direct-Gap Semiconductor,
\prl \textbf{ 105}, 136805 (2010).


\bibitem{Xiao} D. Xiao, G.~B. Liu, W. Feng, X. Xu, and W. Yao, Coupled Spin and Valley Physics in Monolayers of MoS$_2$ and Other Group-VI Dichalcogenides, \prl {\textbf{ 108}}, 196802 (2012).

\bibitem{Mak2013} K.~F. Mak,  K. He, C. Lee, G.~H. Lee, J. Hone, T.~F. Heinz, and J. Shan, Tightly bound trions in monolayer MoS$_2$, Nat. Mater. {\bf
12}, 207 (2013).

\bibitem{Glazov_rmp} G. Wang, A. Chernikov, M.~M. Glazov, T.~F. Heinz, X. Marie, T.
Amand, and B. Urbaszek, Colloquium: Excitons in atomically thin transition metal dichalcogenides, \rmp {\textbf{ 90}}, 021001 (2018).

\bibitem{Rivera} P. Rivera \textit{et al.} Observation of long-lived interlayer excitons in monolayer MoSe$_2$–WSe$_2$ heterostructures, Nat. Commun. {\bf 6}, 6242 (2015).


\bibitem{Latini} S. Latini,  K.~T. Winther, T. Olsen, and K.~S. Thygesen, Interlayer Excitons and Band Alignment in MoS$_2$/hBN/WSe$_2$ van der Waals Heterostructures, Nano
Lett. {\bf 17}, 938 (2017).

\bibitem{Menon} X. Liu, T. Galfsky, Z. Sun, F. Xia, E.-C. Lin, Y.-H. Lee, S. K\'{e}na-Cohen, and V.~M. Menon, Strong light–matter coupling in two-dimensional atomic crystals, Nature  Photonics {\bf 9}, 30 (2015).



\bibitem{BLS} O.~L. Berman, Yu.~E. Lozovik, and D.~W. Snoke, Theory of Bose-Einstein condensation and superfluidity of two-dimensional polaritons in an in-plane harmonic potential, \prb  \textbf{ 77}, 155317 (2008).


\bibitem{BKLS} O. L. Berman, R. Ya. Kezerashvili, Yu. E. Lozovik and D. W. Snoke, Bose–Einstein condensation and superfluidity of trapped polaritons in graphene and quantum wells embedded in a microcavity, {Phil. Trans. R. Soc}. A  \textbf{368}, 5459-5482 (2010).

\bibitem{BKL} O.~L. Berman, R.~Ya. Kezerashvili, and Yu.~E. Lozovik, Spin Hall effect for polaritons in a transition metal dichalcogenide embedded in a microcavity, \prb {\textbf{ 99}}, 085438 (2019).

\bibitem{Cao} Y. Cao \textit{et al.} Unconventional superconductivity in magic-angle graphene superlattices,  Nature {\bf 556}, 43 (2018).

\bibitem{Hunt} B. Hunt \textit{et al.} Massive Dirac fermions and Hofstadter butterfly in a van der Waals heterostructure, Science {\bf 340}, 1427 (2013).

\bibitem{Dean} C.~R. Dean \textit{et al.} Hofstadter's butterfly and the fractal quantum Hall effect in moiré superlattices, Nature {\bf 497}, 598 (2013).

\bibitem{Kim} K. Kim \textit{et al.} Tunable moiré bands and strong correlations in small-twist-angle bilayer graphene, Proc. Natl. Acad. Sci. USA {\bf 114},
3364 (2017).


\bibitem{Yu} H.~Y. Yu, G.~B. Liu, J.~J. Tang, X.~D. Xu, and W. Yao, Moiré excitons: From programmable quantum emitter arrays to spin-orbit–coupled artificial lattices, Sci.~Adv. {\bf 3}, e1701696 (2017).


\bibitem{Wu} F.~C.  Wu, T. Lovorn, and A.~H. MacDonald, Theory of optical absorption by interlayer excitons in transition metal dichalcogenide heterobilayers, \prb {\textbf{ 97}}, 035306 (2018).


\bibitem{Wu_PRL} F.~C.  Wu, T. Lovorn, and A.~H. MacDonald, Topological exciton bands in moiré heterojunctions, \prl  {\textbf{ 118}}, 147401
(2017).


\bibitem{Tran} K. Tran \textit{et al.} Evidence for moiré excitons in van der Waals heterostructures, Nature {\bf 567}, 71 (2019).


\bibitem{Seyler} K.~L. Seyler, P. Rivera, H. Yu, N.~P.
Wilson, E.~L. Ray, D.~G. Mandrus, J. Yan, W. Yao, and  X. Xu, Signatures of moiré-trapped valley excitons in MoSe$_2$/WSe$_2$ heterobilayers, Nature
{\bf 567}, 66 (2019).

\bibitem{Alexeev} E.~M. Alexeev \textit{et al.} Resonantly hybridized excitons in moiré superlattices in van der Waals heterostructures, Nature {\bf 567}, 81 (2019).

\bibitem{Jin} C. Jin et al, Identification of spin, valley and moiré quasi-angular momentum of interlayer excitons, Nat. Phys.
\textbf{ 15} 1140-1144 (2019).




\bibitem{Cavity} S. Dufferwiel \textit{et al.} Exciton–polaritons in van der Waals heterostructures embedded in tunable microcavities, {Nat. Commun.} \textbf{6}, 8579 (2015). 

\end{thebibliography}
\end{document}